# Data-driven Geophysics: from Dictionary Learning to Deep Learning

**Siwei Yu[1] and Jianwei Ma[2,1]**

[1] Center of Geophysics, Artificial Intelligence Laboratory, and School of Mathematics, Harbin Institute of Technology, Harbin, China.

[2] School of Earth and Space Sciences, Peking University, Beijing, China.

Corresponding author: Jianwei Ma (jwm@pku.edu.cn)

**Key Points:**

- A review of state-of-the art artificial intelligence methods in geophysical applications is provided.

- The relationship between traditional dictionary learning methods and deep learning methods is discussed.

- A tutorial for beginners and a discussion of future directions are given.

**Abstract**

Understanding the principles of geophysical phenomena is an essential and challenging task. "Model-driven" approaches have supported the development of geophysics for a long time; however, such methods suffer from the curse of dimensionality and may inaccurately model the subsurface. "Data-driven" techniques may overcome these issues with increasingly available geophysical data. In this article, we review the basic concepts of and recent advances in data-driven approaches from dictionary learning to deep learning in a variety of geophysical scenarios. Explorational geophysics including data processing, inversion and interpretation will be mainly focused. Artificial intelligence applications on geoscience involving deep Earth, earthquake, water resource, atmospheric science, satellite remoe sensing and space sciences are also reviewed. We present a coding tutorial and a summary of tips for beginners and interested geophysical readers to rapidly explore deep learning. Some promising directions are provided for future research involving deep learning in geophysics, such as unsupervised learning, transfer learning, multimodal deep learning, federated learning, uncertainty estimation, and activate learning.

**Plain Language Summary**

With the fast development of artificial intelligence (AI), innumerable students and researchers in the geophysical community would like to know the AI, to learn how to use AI for geophysics, how and what the AI can impact the traditional geophysical methods. We present a review written in the geophysical language, for readers to get the picture for history, recent advances, open problems, and future directions. This review aims to pave the way for more geophysical researchers, students, and teachers to use data-driven AI techniques.

# 1    Introduction

Observation is an important means by which humans come to understand unknown natural phenomena. Geophysics involves the physical mechanics and properties of the earth and space environments at various temporal and spatial scales, and the subject encompasses seismology, gravity fields, magnetic fields, electric fields, the atmosphere, and the internal structures of Earth and other planets. With state-of-the-art observation equipment, the amount of observed data is increasing at an impressive speed. How to process such a large amount of observed data and obtain useful information is a significant problem for further understanding the laws of geophysics. Model- and data-driven methods are two important ways to process geophysical data (Figure 1).

Model-driven geophysics is based on induction and deduction from the perspective of philosophy of science. First, the principles of geophysical phenomena are induced from a large amount of observed data. Mathematical or modeling methods are established based on physical causality and laws, such as the wave equation, the diffusion equation, the heat transfer equation, Maxwell's electromagnetic equation, Newton's law of universal gravitation, Newton's law of motion, and many empirical formulas. Then, the models are used to deduce future or past geophysical phenomena. Model-driven methods have played a vital role in the evolution of geophysical methods. For example, Newton's law of universal gravitation helps find Neptune and Pluto without direct observation at these two planets. Despite the success of model-driven methods, they have limitations in the accurate prediction of complex system states at large spatial and temporal scales, such as in global climate estimation and earthquake prediction. We list several difficult tasks in geophysics in Table 1, in which data-driven methods may be more advantageous than model-driven methods.

To show the bottlenecks of model-driven methods in detail, we use exploration geophysics as an example. Exploration geophysics aims to observe the subsurface of Earth or other planets with physical fields collected at the surface, such as seismic fields and gravity fields. The process of exploration geophysics includes signal processing, modeling, inversion, and interpretation. In the geophysical signal processing stage, the simplest model assumption regarding the shape of underground layers is the linear assumption in small windows (Spitz 1991). Further assumptions include the sparsity (Herrmann and Hennenfent 2008) and low-rank (Oropeza and Sacchi 2011) assumptions, among others. However, the predesigned linear event assumption or sparse transform assumption is not adaptive to seismic data and may lead to low denoising or interpolation quality for data with complex structures. In the seismic modeling and inversion stages, wave equations govern the kinematics and dynamics of seismic wave propagation. Acoustic, elastic, or viscoelastic wave equations introduce an increasing number of factors into the wave equations, and the generated wave field records can precisely estimate real scenarios. However, as the wave equation becomes increasingly complex, the numerical implementation of the equation becomes nontrivial, and the computational cost increases considerably for large-scale scenarios. In seismic interpretation tasks, traditional methods rely on the experience of interpreters, leading to low efficiency and subjective bias. Therefore, an adaptive and automated process in geophysical fields is needed. Figure 2 illustrates model- and data-driven methods in exploration geophysics.

"Data-driven" methods were proposed to overcome the bottlenecks of model-driven approaches. In data-driven geophysics, given a large amount of observed data, the computer first builds a regression or classification model without considering physical causality. This process is also called training. Then, this model performs tasks such as prediction, classification, and recognition on incoming datasets. Dictionary learning (Aharon et al. 2006) and deep learning (DL) (LeCun et al. 2015) are two widely adopted data-driven techniques. In dictionary learning, an adaptive dictionary is learned as a representation of the target data rather than utilizing a predefined dictionary. Dictionary learning is widely applied in seismic data denoising, interpolation, and inversion based on compressive sensing (Candes and Wakin 2008). The key features of dictionary learning are single-level decomposition, unsupervised learning, and linearity. Single-level decomposition means that one dictionary is used to represent a signal. Unsupervised learning means no labels are provided during dictionary learning. In addition, only the target data are used without an extensive training set. Linearity implies that the data decomposition on the dictionary is linear. The above features make the theory of dictionary learning simple; however, they limit the data representation ability of a dictionary. Unlike dictionary learning, DL decomposes data at a deep level and has an excellent representation ability.

DL and artificial intelligence (AI) are very hot topics in recent years. DL is a representative category of AI methods. DL methods train a deep neural network (DNN) through a complex nonlinear mapping process with adjustable parameters based on a large dataset. The trained DNN is used to predict the desired output of the target data for a specific task, such as prediction, detection, and classification. Deep learning encompasses both supervised and unsupervised data-driven approaches depending on whether labels are available. Recently, deep learning methods have been widely adopted in various geophysical applications, such as solid Earth geoscience (Bergen et al. 2019), aftershock pattern analysis (DeVries et al. 2018), and Earth system analysis (Reichstein et al. 2019).

A review article about DL in solid Earth geoscience was recently published in Science (Bergen et al. 2019). The topic includes a variety of machine learning techniques, from traditional methods, such as logistic regression, support vector machines, random forests and neural networks, to modern methods, such as deep neural networks and deep generative models. The article stresses that machine learning will play a key role in accelerating the understanding of the complex, interacting and multiscale processes of Earth's behavior. Since 2019, AI geophysics has made rapid progress. Our review will introduce AI geophysics at the leading edge, mainly focusing on DL, will cover a variety of geophysical applications, from deep to the earth's core and far into outer space, and will mainly focus on exploration geophysics. Furthermore, this review will help readers transfer existing knowledge on dictionary learning and compressive sensing to DL. This review intends not only to provide a glance at the most recent DL research related to geophysics for geophysical readers but also to provide a cookbook for beginners who are interested in DL, from geophysical students to young researchers.

This review is organized as follows. First, the background, target, and outline of this review are given in the introduction (S1). The review part consists of three layers. The first layer contains concepts, and we introduce the basic idea of dictionary learning and DL (S2). The second layer contains detailed techniques (S3, S4). Applications in exploration geophysics are introduced following each method to better explain the concepts of dictionary learning and DL. The third layer presents AI applications in other geophysical areas (S5). The relationship between traditional methods and DL is further discussed in S6. A tutorial section for beginners (S7) and a discussion of future directions (S8) are given as extensions of this review. S9 summarizes this review.

## 2 General theory

Readers who are already familiar with general theory in dictionary learning, compressive sensing and DL may wish to skip to Section 3. We denote scalars by italic letters, vectors by bold lowercase letters and matrices by bold uppercase letters. In geophysics, the goal is to invert unknown parameters $\mathbf{x}$ from an available dataset $\mathbf{y}=\mathbf{Lx}$. $\mathbf{L}$ is a forward or degraded operator in geophysical applications, such as denoising, reconstruction, or full-waveform inversion. However, $\mathbf{L}$ is usually ill-conditioned or not invertible. Compressive sensing approximates $\mathbf{x}$ by forming an optimization objective loss function $E(\text{x})$ with an additional constraint $R$:

$$E(\mathbf{x}) = D(\mathbf{Lx}, \mathbf{y}) + R(\mathbf{x})$$

where $D$ is a similarity measurement function. Typically, the L2-norm $\|\mathbf{Lx} - \mathbf{y}\|_2$ is used for smooth measurement. $R$ is a regularization term. Sparsity is a popular regularization term adopted in compressive sensing, where $R(\mathbf{x}) = \|\mathbf{Wx}\|_1$. $\mathbf{W}$ is a sparse transform with several vectorized basis. $\mathbf{W}$ is also termed as dictionary. The goal of dictionary learning is to train an optimized sparse transform $\mathbf{W}$, which is used for the sparse representation of $\mathbf{x}$. Dictionary learning involves learning $\mathbf{W}$ via matrix decomposition with constraints $R_w$ and $R_c$ on the dictionary $\mathbf{W}$ and coefficient $\mathbf{v}$.

$$E(\mathbf{W}, \mathbf{v}) = D(\mathbf{W}^{\mathrm{T}}\mathbf{v}, \mathbf{x}) + R_w(\mathbf{W}) + R_c(\mathbf{v}) \qquad (1)$$

Unlike dictionary learning, deep learning treats geophysical problems as regression or classification problems. A DNN $F$ is used to approximate $\mathbf{x}$,

$$\mathbf{x} = F(\mathbf{y}; \boldsymbol{\Theta})$$

where $\boldsymbol{\Theta}$ is the parameter set of the DNN. From the view of mathematics, a DNN provides a high-dimensional and nonlinear mapping from $\mathbf{y}$ to $\mathbf{x}$. From the perspective of biology, the architecture of a DNN is bionic and includes biological nerve cells in a multilayer structure. Each layer contains several neural units, with input, output, and nonlinear activation features that are analogous to axons, dendrites, and the cell unit. The input and output of a neuron are connected to the neurons in the neighboring layers, and $\boldsymbol{\Theta}$ represents the connection weights. A particular connection format, the convolutional filter, is often used in modern DNNs to share the parameters among different neurons.

Deep learning aims to build a high-dimension approximation between two sets $\mathbf{X} = \{\mathbf{x}_i, i = 1 \cdots N\}$ and $\mathbf{Y} = \{\mathbf{y}_i, i = 1 \cdots N\}$, i.e., the inputs and labels, with a DNN. The approximation is achieved by minimizing the following loss function to obtain an optimized $\mathbf{\Theta}$:

$$E(\mathbf{\Theta}; \mathbf{X}, \mathbf{Y}) = \sum_{i=1}^{N} \left\| \mathbf{x}_i - F(\mathbf{y}_i; \mathbf{\Theta}) \right\|_2^2$$

If $F$ is differentiable, a gradient-based method can be used to optimize $\mathbf{\Theta}$. However, a large Jacobi matrix is involved when calculating $\nabla_{\mathbf{\Theta}} E$, making it infeasible for large-scale datasets. A back-propagation method (LeCun et al. 1997) is proposed to compute $\nabla_{\mathbf{\Theta}} E$ and avoid calculating the Jacobi matrix. In the following sections, we first introduce dictionary learning from the traditional K-means method and the widely used K-SVD approach; a recent fast algorithm with a tight data-driven framework and other deep learning methods are also presented. Each subject is introduced along with applications in exploration geophysics. A mapping diagram of the primary references used in this paper is given in Figure 3.

## 3    Dictionary learning

### 3.1    K-means

Clustering is used to group geophysical attributes into several classifications. For example, we need to decide whether a region contains fluvial facies or faults based on stacked sections. K-means (Hartigan and Wong 1979), which is a classical clustering algorithm, can be treated as a dictionary learning method with an extremely sparse representation, where only one dictionary component is allowed, and the representation coefficient must be one. Though simple, we can briefly explore the basic steps in dictionary learning with this approach.

K-means aims to cluster $N$ given samples with $M$ features into $K$ groups. K-means applies two steps per iteration with $K$ randomly initialized cluster centers. i) Assign the training samples to the nearest cluster center. ii) Update each cluster center based on the weighted center of the attached samples. Figure 4 shows an example of how K-means splits a dataset into two classes based on two selected features.

K-means, and the corresponding improved methods, is used for signal classification in geophysics. Because K-means is sensitive to feature selection, Galvis et al. (2017) suggested that seismic attributes be selected based on the notion of similarity and that these attributes can be used in the classification of surface waves with the K-means approach. K-means is also sensitive to outliers. Song et al. (2018) propose an adaptive-phase K-means method. The advantage of using the phase distance as a similarity measure is that it provides robustness in the presence of horizon error. The classification result of K-means depends on the user-specified number of clusters. Waheed et al. (2019) showed that the density-based spatial clustering method does not require a specification for the number of clusters and reduces the cost of automatic velocity selection compared to that in the traditional K-means approach. De Lima and Marfurt (2018) proposed a combination of principal component analysis and K-means for the classification of airborne gamma ray spectrometry.

## 3.2 K-SVD

Similar to K-means, most dictionary learning algorithms consist of two steps: i) sparse coding and ii) dictionary updating. Unlike K-means, the sparse representation is not limited to one component, and the number of representation coefficients is also not limited. The method of optimal directions (MOD) (Engan et al. 2002) uses orthogonal matching pursuit for sparse coding and a second-order Newtonian method for dictionary updating. Beckouche and Ma (2014) use the MOD method for dictionary learning and sparse approximation in seismic denoising. The dictionary updating approach in MOD has favorable flexibility and simplicity; however, it is relatively impractical for large dictionaries since matrix inversion is involved.

K-SVD (where SVD is singular value decomposition) (Aharon et al. 2006) shares the same sparse coding structure as MOD, but several improvements in dictionary updating are given. First, one component of the dictionary is updated at a given time, and the remaining terms are fixed. Second, a rank-1 approximation SVD algorithm is used to obtain the updated dictionary and coefficients simultaneously, thereby accelerating convergence and reducing computational memory use compared to those in the MOD. K-SVD is applied in geophysics with preferred extensions. Nazari Siahsar et al. (2017) split the training data into different slices and trained different dictionaries with a shared sparse coefficient matrix. Such a strategy allows 3-D datasets to processed with a reasonable time cost for all training patches.

### 3.3 Data-driven tight frame

Despite the success of K-SVD in signal enhancement and compression, dictionary updating is still time consuming in regard to high-dimensional and large-scale datasets, such as 3-D prestacked data in seismic exploration. K-SVD includes one SVD step to update one dictionary term. Can the entire dictionary be updated by one SVD for efficient improvement? Cai et al. (2014) proposed a data-driven tight frame (DDTF) by enforcing a tight frame constraint on the dictionary. The tight frame condition is a slightly weaker condition than orthogonality, for which the perfect reconstruction property holds. With the tight frame property, dictionary updating in DDTF is achieved with one SVD, which is hundreds of times faster than K-SVD.

Liang et al. (2014) first utilized 2-D DDTF in seismic data interpolation. The extension of dictionary learning to high dimensions is straightforward since the data are vectorized at the patch scale. Patches are blocks generated from original data division into training samples. Yu et al. (2015) extend DDTF to 3-D and 5-D with applications in seismic data interpolation. The training patches for dictionary learning are a random subset of all patches. An example of a learned dictionary with 3-D DDTF for a seismic volume is shown in Figure 5. Yu et al. (2016) designed a Monte Carlo selection method based on a training set. The patches with high variance were selected with high probability to further improve efficiency. Liu et al. (2017) proposed tensor DDTF, in which high-dimensional data are obtained by tensor products, to save computational resources and constrain data structures. Liu et al. (2018) and Liu and Ma (2019) proposed graph DDTF, in which a binary tree is used to cluster training patches. DDTF is implemented for each cluster to obtain a sparse dictionary with similar patches as the original dictionary. Wang and Ma (2019) and Wang et al. (2019) proposed adaptive DDTF and group-sparsity DDTF for the preservation of weak signals by considering the similarity among different patches.

## 4    From dictionary learning to deep learning

Though both are data-driven methods, deep learning differs from dictionary learning in three aspects: the depth of decomposition, the amount of training data, and the nonlinear operators. Dictionary learning is usually a single-level matrix decomposition problem. Rubinstein et al. (2010) proposed double sparsity (DS) dictionary learning to explore deep decomposition. The motivation of DS is that the learned dictionary atoms still share some underlying sparse pattern for a generic dictionary. In other words, the dictionary is represented with a sparse coefficient matrix multiplied by a fixed dictionary, as in discrete cosine transform. Inspired by DS dictionary learning, can we propose triple, quadruple or even centuple dictionary learning? We know cascading linear operators are equivalent to a single linear operator. Therefore, using more than one fixed dictionary does not improve the signal representation ability compared to that ability of one fixed dictionary if no additional constraints are provided. In deep learning, nonlinear operators are combined in such a deep structure. A neural network with one hidden layer and nonlinear operators can represent any complex function with a sufficient number of hidden neurons. To fit a neural network with many hidden neurons, we need an extensive training set, while dictionary learning involves only one target data. A comparison of the learned features in dictionary learning and deep learning is shown in Figure 6.

Deep learning is a machine learning method based on a DNN, which is an artificial neural network (ANN) with many layers. ANNs are widely in machine learning and date from the late 1940s. In a multilayer perceptron (MLP), a specific type of ANN, neurons are organized into different groups called layers. Neurons in adjacent layers are connected by connection weights. The output of a neuron is the weighted summation of the output of the neurons from the previous layers, and each output is then input into a nonlinear activation function. The simplest MLP includes an input layer, a hidden layer, and an output layer.

Poulton (2002) published a review article on ANN methods in geophysics in 2000. Since 2000, many pioneers have applied machine learning methods in geophysics, and deep learning has slowly become popular. Limited by the length of this review, we only recall some such studies. Lim (2005) characterized reservoir properties using fuzzy logic and an ANN for well data. Huang et al. (2006) explored seismic data parameter determination and pattern detection with an ANN. Helmy et al. (2010) applied hybrid computational models to characterize oil and gas reservoirs. Zhang et al. (2014) proposed using a kernel-regularized least-squares (Evgeniou et al. 2000) method for fault detection from seismic records. Jia and Ma (2017) suggest using supported vector regression (Cortes and Vapnik 1995) for seismic interpolation. The authors used linearly interpolated data and original data as the inputs and outputs, respectively, and supported vector regression was used to obtain the relation between the inputs and outputs. They claimed that no assumptions were imposed on the data and that no parameter tuning was required for interpolation.

Our review will focus on DNNs. The number of layers in an ANN has a significant effect on the fitting and generalization abilities of an algorithm. Early ANNs were restricted to a few layers due to the computational capacity of the available hardware. With the development of hardware and optimization algorithms, ANNs are expanding towards deep layers. However, MLPs encounter computational and storage problems due to the massive number of parameters required when the network deepens or the size of the inputs increases in practical applications. In addition, an MLP requires preselected features as inputs into the neural network and ignores the structure of the input entirely, with full reliance on experience. Qi et al. (2020) proposed feature selection for machine learning facies analysis. An exhaustive search method, which took 638 hours for the candidate attributes, was used to determine both the optimal number and combination of parameters. To reduce the number of parameters in an MLP, convolutional neural networks (CNN) (McCann et al. 2017) were proposed to share parameters with convolutional filters.

## 4.1 Convolutional neural networks

CNNs were proposed to consider local coherency and reduce the number of weight parameters. CNN uses convolutional filters to restrict the inputs of a neural network to within a local range. The convolutional filters are shared by different neurons in the same layer. A CNN uses original data rather than selected features as an input set. Pooling layers are used in CNNs to extract key features by subsampling the input set. CNNs have developed rapidly since 2010 for image classification and segmentation, and some popular CNNs include VGGNet (Simonyan and Zisserman 2015) and AlexNet (Krizhevsky et al. 2017). CNNs are also used in image denoising (Zhang et al. 2017) and super-resolution tasks (Dong et al. 2014). The above CNNs are named vanilla CNNs, which are CNNs with simple sequential structures. Vanilla CNNs are used for regression and classification tasks. In regression tasks, the outputs are continuous variables; in classification tasks, the outputs are discrete variables. Compared to K-means, deep learning achieves complex classification with a multilayer feature extractor and simple classifiers.

Additional deep learning network architectures have been proposed for specific tasks based on MLPs or vanilla CNNs (Figure 7a,b). An autoencoder (AE) is a network in which the inputs and outputs are the same. Hidden layers in AEs extract deep features that are typically used for unsupervised classification with the help of K-means. A deep convolutional autoencoder (CAE, Figure 7c) is an AE with convolutional layers that acts as a feature extractor. U-Net (Ronneberger et al. 2015) (Figure 7d) uses skip connections to bring low-level features to a high level. The generative adversarial network (Goodfellow et al. 2014, Creswell et al. 2018) (GAN, Figure 7e) aims to reproduce data examples with the same distribution as the training set. A GAN contains a generative network and a discriminative network. The generative network tries to produce a nearly real image. The discriminative network tries to distinguish whether the input image is real or generated. Therefore, such a game will finally allow the generative network to produce fake images that the discriminative network cannot distinguish from real images. CycleGAN (Zhu et al. 2017) is a GAN with two generative networks and two discriminative networks, such that a cycle mapping between two datasets is trained. Recurrent neural networks (RNNs, Figure 7f) are commonly used for prediction tasks based on sequential data, and the prediction for the current input depends on the history of inputs fed into the neural network. Long short-term memory (LSTM) (Hochreiter and Schmidhuber 1997) is a widely used RNN that considers how much historical information is forgotten or remembered. We introduce the concepts and applications of the above DNNs in detail in the following sections.

## 4.2 Vanilla convolutional neural networks

Vanilla CNNs are the most popular CNNs if many training samples and labels are available. Cascading convolutional layers, nonlinear layers, and data regularization layers provide a remarkable fit for the training samples in regression tasks. Pooling layers are used for feature extraction in classification tasks. Vanilla CNNs are reliable for most applications in geophysics, such as denoising, interpolation, velocity modeling, and data interpretation.

In the seismic denoising area, Yu et al. (2019) proposed a denoising CNN (DnCNN) (Zhang, Zuo et al. 2017) based method for three kinds of seismic noise. The DnCNN developed was composed of convolutional, batch normalization, and rectified linear unit layers. In this approach, the final output is equal to the network output plus the input, which is called residual learning, i.e., the output of the network represents noise. The concept of residual learning is similar to that used in a residual network (He et al. 2016) to avoid vanishing gradients. Random and linear noise are manually added to synthetic datasets, and multiple data sets are generated with the acoustic wave equation. In this case, transfer learning (Donahue et al. 2014) is used for field data denoising. Different kinds of noise are processed with the same network architecture but different training sets. An example of scattered ground-roll attenuation is shown in Figure 8. Scattered ground roll is mainly observed in desert area, caused by the scattering of ground roll when the near surface is laterally heterogeneous. Scattered ground roll is difficult to remove because it occupies the same F-K domain as reflected signals. DnCNN was used to remove scattered ground roll successfully. Wu et al. (2019) claimed that in a traditional CNN, the labels of clean data are difficult to obtain. They used multiple trials involving user-generated white noise to simulate real white noise. Additionally, the inputs were decomposed with the variational mode decomposition method (Dragomiretskiy and Zosso 2014) to obtain a few modes with different frequency supports, which were then fed into a CNN.

In the seismic interpolation field, Wang et al. (2019) proposed the use of ResNet for the reconstruction of regularly missing data. The training set consisted of synthetic and field samples. The input of the network was preprocessed with a bicubic interpolation algorithm. Zhang et al. (2020) trained a denoised neural network with a natural image dataset and used the trained network in the project onto a convex set (Abma and Kabir 2006) framework for seismic data interpolation. Therefore, no new networks were required for the interpolation of other datasets or other tasks. Figure 9 gives the training set and a simple interpolation result (Zhang, Yang et al. 2020).

In seismic deblending, Zu et al. (2020) constructed an end-to-end deblending CNN. The trained network was iteratively applied to blended data. The authors claimed that networks trained with both synthetic and field datasets perform well with real input datasets. Sun et al. (2020) also used an end-to-end CNN for deblending. Different hyper-parameters, such as the number of layers, number of filters, and size of the filters, were used to construct a network that was optimal for seismic data. They used field datasets to construct the training set, which inevitably contained some noise contamination associated with the labels. Nakayama et al. (2019) presented a method for designing acquisition parameters, including blending, source, and receiver positions, based on a genetic algorithm and CNN. The genetic algorithm was used to produce combinations of acquisition parameters, and the CNN was used to classify the combinations. Finally, a deblending algorithm was used to obtain a clean signal. If this signal was close to the original signal, the iteration stopped; if not, the algorithm proceeded with new parameters.

In velocity analysis and inversion, Araya-Polo et al. (2018) used a CNN for seismic tomography and obtained a promising result for synthetic 2D data. Wang and Ma (2020) proposed a network combined with fully connected layers and fully convolutional layers (FCN). The FCN used a contracting encoder and an expansive decoder corresponding to feature extraction and function fitting, respectively. The FCN yielded outputs with the same size as the inputs. The input was a seismology data set with a cross-well geometry, and the output was a velocity model. Notably, the authors used smoothed natural images as seismic models, thus producing a large number of models to construct the training set. Figure 10 shows how Wang and Ma (2020) converted a three-channel color image to a velocity model. Park and Sacchi (2019) developed a CNN to directly estimate stacking velocities. The portions with different time slices as channels were used as inputs. The root square velocity was the output. For a much different dataset, the authors used transfer learning instead of network training from random initialization.

Ovcharenko et al. (2019) proposed a DL framework for extrapolating the frequency range of seismic data from high to low frequencies. The inputs and outputs of the network were multiple high-frequency and single low-frequency representations of a shot gather. Moreover, 0.25 Hz data were obtained from 2 to 4.5 Hz frequencies. Low-frequency information was used for the initialization of full-waveform inversion (FWI).

In the attribute inversion area, Das et al. (2019) proposed a 1-D CNN for seismic impedance inversion. The training set consisted of synthetic datasets and contained six output features, such as the spherical variogram ranges of facies, phases, and the central frequency. The uncertainty was computed with an approximate Bayesian computational method. You et al. (2020) predicted anisotropy information from conventional well logs based on a DNN, and the method was generalized for use with field data.

### 4.3    Convolutional autoencoder

A CAE is a type of CNN consisting of an encoder and a decoder. The encoder uses convolutional layers and pooling layers to extract critical features in a latent space from the inputs, resulting in a contracting path. The decoder uses deconvolutional layers and unpooling layers to decode the features into the original data space, resulting in an expanding path. A CAE works in both supervised and unsupervised ways. If labels are provided as outputs, a CAE is a supervised regression network. If the outputs are the same as the inputs, a CAE works in an unsupervised way, and the latent features are used for other tasks, such as clustering. The learned latent features can also be used for dimension reduction in large-scale tasks.

In seismic data processing, Wang et al. (2020) proposed a CAE-based interpolation method for irregular sampling. The subsampled dataset is the input, and the complete dataset is the output. Transfer learning is used when the method is applied to field data. Wu et al. (2019) treated first-arrival selection as an image segmentation problem with a CAE. Anything prior to the first arrival is set to zero, and all instances after the first arrival are set to one. This method works well for noisy situations and field datasets. For the training set, a subset of traces generated with a simple model is used. Gao et al. (2019) used CAE for dimension reduction in FWI to estimate longwave information, where the latent parameters were optimized instead of the whole dataset.

In attribute analysis, Duan et al. (2019) used a CAE to extract the features of 1D data and K-means for clustering. They used Kullback-Leibler divergence to measure the similarities between the two distributions. He et al. (2018) and Qian et al. (2018) used an AE to extract seismic features in an unsupervised way, and then a K-means clustering method was used to classify the seismic facies. Qian, Yin et al. (2018) built a physical model and used field data from the Liziba survey to test the proposed method.

## 4.4 U-Net

U-Nets have U-shaped structures and skip connections. The skip connections bring low-level features to high levels. U-Net was first proposed for image segmentation and has been applied in seismic data processing, inversion, and interpretation. The U-structure with a contracting path and expanding path makes every data point in the output contain all information from the input, such that the approach is suitable for mapping data in different domains, such as inverting velocity from seismic records in FWI. The input size of the test set must be the same as that in the training set for a trained U-Net.

In seismic data processing and inversion, Mandelli et al. (2018) used a U-Net for the interpolation of seismic data, and prestack images with and without missing traces were used as inputs and outputs. Hu et al. (2019) proposed a U-Net-based first-arrival selection method by formulating a binary segmentation problem. Yang and Ma (2019) proposed a new general velocity model construction method based on U-Net. The inputs were seismology data sets generated by the acoustic wave equation from surface survey, and labels were the velocity models. This method is useful for generating low-frequency models for the initialization of traditional FWI. Low-frequency information helps FWI converge. Figure 11 shows the velocity prediction results from Yang and Ma (2019). Unlike the conventional inversion method based on physical models, supervised deep learning methods are based on big-data training rather than prior-knowledge assumptions. Unlike in FWI, after network training is completed, the reconstruction costs are negligible. Moreover, little human intervention is needed, no initial velocities are involved, and no cycle-skipping problem exists. Instead of regression from seismic records to velocity models, Zhang and Alkhalifah (2019) used deep learning to estimate the distribution of facies from the results of conventional FWI, and the facies were used to constrain a new iteration of the FWI approach several times. In this method, deep learning was integrated into FWI as part of a model constraint.

In seismic interpretation, Wu et al. (2020) built an approximately realistic 3-D training dataset by randomly choosing folding and faulting parameters in a reasonable range. Then, the dataset was used to train a 3D U-Net for seismic structural interpretation for features such as faults, layers, and dips in field datasets. Building realistic synthetic datasets rather than handcrafted field datasets is more efficient and can produce similar results. Wu et al. (2019) used an end-to-end U-Net for 3D seismic fault segmentation. The inputs were seismic images, and the outputs were ones, indicating faults, and zeros, indicating nonfaults. A class-balanced binary cross-entropy loss function was used to adjust the data imbalance so that the network was not trained to predict only zeros. A network trained only on synthetic data worked well when field datasets were considered. Wu et al. (2019) treated the horizon interpretation problem as an image classification task; they used U-Net and post-stack traces located on a user-defined coarse grid as the inputs and manually picked horizons as labels. This approach is semi-automated because the horizons must be labeled for the coarse grid. The traces were processed individually. An example of synthetic post-stack image and field data fault analysis is shown in Figure 12, as published by Wu, Geng et al. (2020).

It is convenient to apply U-Net in various geophysical applications. Here, we give two examples: velocity picking and first-arrival picking. Figure 13 shows the results of using U-Net for velocity picking. The inputs are seismological data, and the outputs are ones where the picks are located and zeros elsewhere. Figure 14 shows the results of the phase picking based on U-Net. We used 8000 synthetic samples. A gradient constraint was added in the loss function to enhance the continuity of the picked positions. Seismological data sets were used as inputs. For the output, three classifications were set: zeros above the first arrival, ones below the first arrival, and twos for the first arrival. The training dataset was contaminated with strong noise and had missing traces. The predicted picking were close to the labels. First-arrival picking based on deep learning has been applied for realistic seismic data processing by using different neural networks (Hu, Zheng et al. 2019, Wu, Zhang et al. 2019).

### 4.5 Generative adversarial networks

GANs can be applied in adversarial training for two CNNs: one generator to produce a fake image and one discriminator to distinguish the produced image from real images. When training the discriminator, the real dataset and generated dataset correspond to labels one and zero, respectively. Additionally, when the generator is trained, all datasets correspond to the label one. A GAN is used to generate samples with similar distributions as the training set. The generated samples are used for simulating realistic scenarios or expanding the training set. Zhu, Park et al. (2017) proposed an extended GAN, named CycleGAN, with two generators and two discriminators for signal processing. In CycleGAN, a two-way mapping is trained for mapping two datasets from one to the other. The training set CycleGAN is not necessarily paired, as in a vanilla CNN, which makes it relatively easy to construct training sets in geophysical applications with few labels.

To artificially expand labeled data sets, Wang et al. (2019) proposed the GAN-based model EarthquakeGen. The detection accuracy was greatly improved by performing artificial sampling for the training set. Si et al. (2020) proposed the use of CycleGAN for ground-roll attenuation. The training set consisted of synthetic and field-derived seismic data. Zhang et al. (2019) proposed a seismic enhancement algorithm based on a GAN with time resolution improvements. Lipari et al. (2018) used a GAN to map low-quality migrated images to high-quality images and the corresponding reflectivity images. Siahkoohi et al. (2019) proposed a GAN to produce a high-quality wavefield from a low-quality wavefield in the context of surface-related multiples, ghosts, and dispersion. The training procedure consisted of initial training and transfer learning. The initial training was performed with datasets from nearby surveys. Transfer learning was performed with a small training set with high-fidelity data from the current dataset. The GAN used did not require training set pairing. Only two sets with and without high fidelity are needed. Wang et al. (2019) proposed a 1D CycleGAN-based impedance inversion algorithm to mitigate the dependence of vanilla CNNs on the amount of labeled seismic data available.

### 4.6   Recurrent neural networks

In time-sequenced data processing applications, RNNs use the output of a network as the input of the subsequent process to consider the historical influence. RNNs are used for the prediction of new outputs from a sequential input, such as predicting new words from an input sentence. The prediction accuracy of LSTM increases with the amount of historical information considered. In geophysical applications, RNNs are used for predicting the next sample of a time-sequenced or spatially sequenced dataset. RNNs are also used for wavefield simulation based on a time-dependent network form.

In seismic data processing, Payani et al. (2019) used an RNN to estimate the relationships among samples in a seismic trace; they found that 16 bits are needed for lossless representation instead of 32 bits per sample. Chen et al. (2020) applied LSTM for the denoising of magnetotelluric data with prediction data samples in a trace. Li et al. (2019) utilized an RNN to consider the spatial continuity and similarity of adjacent traces in facies analysis.

In seismic modeling and inversion, Sun et al. (2020) constructed an RNN for wave modeling and inversion, and the network parameters corresponded to the selected velocity model. The structure of an RNN is similar to finite different time evolution. Therefore, optimizing an RNN is equivalent to seismic waveform inversion. Experiments with various optimization algorithms, including gradient descent, conjugate gradient, adaptive moment, and limited-memory Broyden-Fletcher-Goldfarb-Shanno algorithms, have been performed. The results have indicated that first-order methods perform better than second-order methods. Liu (2020) extends RNN for simultaneous inversion of velocity and density. Figure 15 shows the structure of a modified RNN based on the acoustic wave equation used in Liu (2020). The diagram represents the discretized wave equation with a flow chart implemented in an RNN. The inversion process in full waveform inversion is the training process of RNN. Fabien-Ouellet and Sarkar (2019) combined three networks: a CNN with a CMP gather input and a semblance output, an RNN for data reduction, and LSTM for velocity decoding. This design was inspired by the information flow in semblance analysis. The proposed method works well for 1D layered velocity models. We give an example of using an RNN for simultaneous velocity and density inversion.

## 5  AI geophysical applications

We investigate more AI geophysical applications apart from exploration geophysics in this section. The topics are roughly arranged by the order from the earth's core to outer space.

### 5.1  The earth's core, mantle and crust

Understanding the interior of the earth is a challenging task since observations are mainly limited on the earth's surface. The earth is roughly divided into core, mantle and crustal layers from inside to the surface; however, the detailed structures and properties of the deep side of the earth are not clear. Kim et al. (2020) discovered a previously unrecognized ultra-low-velocity zone at the core-mantle boundary beneath the Marquesas Islands with a manifold learning method called the Sequencer. Thousands of diffracted waveforms are arranged in a sequential order on the manifold, where half reveal loud signals indicating the ultra-low-velocity zone (Figure 16). Shahnas et al. (2018) predicted mantle flow processes by employing support vector machines and training samples from numerical convection models. Shahnas and Pysklywec (2020) used deep learning to predict the mantle thermal state of simplified model planets. They achieved an accuracy of 99% for both the mean mantle temperature and the mean surface heat flux compared to the calculated values. They claimed that deep learning can be employed in more complex mantle states. Cheng et al. (2019) used DNN to map Rayleigh surface wave velocities to crustal thickness in eastern Tibet and the western Yangtze craton. The training dataset is generated based on seismic wave forward modeling theories.

The earth is covered by soil and rocks. Fang et al. (2017) used LSTM to predict historical soil moisture with high fidelity from two recent years of satellite data, showing LSTM's potential for hindcasting, data assimilation, and weather forecasting. Anantrasirichai et al. (2018) used a CNN to classify interferometric fringes in wrapped interferograms with no atmospheric corrections for detecting volcanic deformation.

### 5.2  Earthquake

The goal of earthquake data processing is quite different from that of exploration geophysics; therefore, this section focuses on deep learning-based earthquake signal processing.

The preliminary processing of earthquake signals includes classification to distinguish real earthquakes from noise and arrival picking to identify the arrival times of primary and secondary waves. Further applications involve earthquake early warning (EEW) analysis and Earth tomography. Deep learning has shown promising results in these applications.

### 5.2.1 Classification

Meier et al. (2019) trained five DNNs for seismic signal and noise discrimination. The training set contained 374,000 earthquake and 946,000 noise records from three channels. Li et al. (2018) trained machine learning algorithms on an extensive dataset to discriminate earthquake P waves from local impulsive noise. Linville et al. (2019) used an RNN and a CNN to identify events as either quarry blasts or earthquakes. The purpose of volcano seismic detection is similar to that of EEW, i.e., determining whether an event is dangerous. Malfante et al. (2018) extracted 102 features from acoustic and seismic fields. Six classes were considered: long-period events, volcanic tremors, volcano-tectonic events, explosions, hybrid events, and tornados.

We provide an example of using the wavelet scattering transform (WST) (Mallat 2012) and a support vector machine for earthquake classification with a limited number of training samples. The WST involves a cascade of wavelet transforms, a module operator, and an averaging operator, corresponding to convolutional filters, a nonlinear operator, and a pooling operator, respectively, in a CNN. The critical difference between the WST and a CNN is that the filters are predesigned with the wavelet transform in the WST. In our case, only 100 records were used for training, and 2000 records were used for testing. We obtained a classification accuracy as high as 93% with the WST method. Figure 17 shows the architecture of the WST algorithm.

### 5.2.2 Arrival picking

Wang et al. (2019) proposed PickNet to choose natural seismic arrivals based on a deep residual network. The selected arrivals were used for seismic tomography, and the underground structure of Japan was reconstructed. Ross et al. (2018) trained a CNN for arrival picking and polarity classification. The training set contained 19.4 million seismograms. They achieved remarkably high picking and classification accuracies close to or better than those obtained by human experts. Zhao et al. (2019) proposed using U-Net for P- and S-phase arrival picking and achieved superior results compared to those of the short-time average over the long-time average (STA/LTA) method. Zhou et al. (2019) developed a hybrid event detection and phase-picking algorithm with both CNNs and RNNs.

### 5.2.3 Earthquake early warning, prediction, and Earth tomography

Zhang et al. (2020) used a CNN to locate seismic sources from received waveforms at several stations. This method worked well for small earthquakes ($M_L$<3.0) with low SNRs, for which traditional methods fail. The prediction results and errors of earthquake source locations are indicated in Figure 18. Ross et al. (2019) generated millions of synthetic sequences to train the PhaseLink network. Associating seismic phases involves classifying seismological records from the same source into one set. This concept is similar to seismic registration. Zhang and Curtis (2020) used variational inference for seismic tomography. This method obtains the mean and variance as outputs. Yamaga and Mitsui (2019) analyzed the relationship between a strong earthquake and postseismic deformation. The dataset was obtained with the Global Navigation Satellite System and was relatively small, with 153 training points and 38 testing points. An RNN was used to learn the corresponding relationships, and the results were far more accurate than those of traditional regression methods. Rouet-Leduc et al. (2017) used a random forest model to predict earthquakes in the laboratory. They used acoustical signals as inputs and predicted laboratory fault failures. The machine learning method identified a signal emitted from the fault zone previously thought to be low-amplitude noise.

## 5.3   Water resources

Water on Earth has a great impact on ecosystems and natural disasters. The oceans contain most of the water on Earth. Estimating the dynamic parameters of oceans is a challenging and important task. Machine learning can predict ocean parameters, such as subsurface temperatures (Hua et al. 2018) and oxygen levels (Giglio et al. 2018). Hua et al. (2018) retrieved anomaly temperatures on the subsurface of the ocean from satellite observations based on the random forest method. They used surface height, temperature, salinity, and wind as inputs. Giglio et al. (2018) estimated the oxygen level in the Southern Ocean. The authors used temperature and salinity as the input for the random forest model. Wang et al. (2019) used LSTM to predict the Loop Current in the Gulf of Mexico. They predicted the Loop Current evolution within 40 kilometers nine weeks in advance.

The sea level is increasing due to global warming. In summer, the meltwater on Arctic ice will abord more solar radiation and accelerate  ice melt. Wright and Polashenski (2020) measured the percentage of pond coverage from Arctic sea ice based on a random forest model. They use low-resolution satellite imagery as input to cover a larger spatial range. Similarly, Barbat et al. (2019) estimated the size of icebergs in the pan-Antarctic near-coastal zone. To cover the whole Antarctic continent, they used a machine learning method that operates on low-resolution synthetic aperture radar (SAR) imagery. Liu et al. (2019) used U-Net to predict coastal inundation mapping from synthetic aperture radar imagery information, providing a better understanding of the geospatial and temporal characteristics of coastal flooding.

In addition to oceans, water is stored in different forms, such as rivers, lakes, rain and snow. Sun et al. (2019) used a CNN to estimate groundwater storage in India. The CNN was trained to compensate for missing components between the satellite data and NOAH-simulated data. Once trained, the CNN was used to correct the NOAH-simulated data without using the satellite data. Limited by the resolution of the remote sensing imagery, Ling et al. (2019) measured river widths on a subpixel scale. With a superresolution CNN, they obtained highly accurate results. Guillon et al. (2020) predicted river channel types from geospatial data in a large river basin using a random forest model (Figure 19). They operated on a coarse scale to cover a large area and achieved an accuracy of 61% for ten types of channels. Read et al. (2019) used an LSTM-based machine learning method to predict the temperature of lake water. They achieved an RMSE reduction of 0.5 ℃ relative to traditional methods.

Akbari Asanjan et al. (2018) also proposed an LSTM-based learning method to predict rainfall runoff. They predicted hourly runoff for a twenty-four-hour period using observations such as those of rainfall and runoff. Tang et al. (2018) used a DNN to estimate rain and snow at high latitudes. They used passive microwave data, infrared data and environmental data from spaceborne satellite radar data to improve precipitation predictions. A review paper of water science based on deep learning is provided in Shen (2018).

### 5.4 Climatology and atmospheric science

Atmospheric science observes and predicts climate, weather and atmospheric phenomena. A complete observation of global atmospheric parameters is difficult since the earth is extremely large and sensor locations are limited. Kadow et al. (2020) chose a CNN-based inpainting algorithm to reconstruct missing values in global climate datasets such as HadCRUT4 (Figure 20). Minnis et al. (2016) used a neural network to estimate the optical depth of ice clouds for the prediction of future water paths. Multispectral infrared radiances were used to reduce the dependence on solar illumination conditions.

Precipitation observed from satellites is on a coarse scale, limiting its resolution in small regions. Sharifi et al. (2019) proposed a downscaling algorithm based on machine learning to map satellite observations to a fine spatial resolution. Whitburn et al. (2016) estimated the global $NH_3$ level from measurements of an infrared atmospheric sounding interferometer. A neural network was used to convert the spectral hyperspectral range index to the $NH_3$ level. Xu et al. (2018) proposed using machine learning to reconstruct evapotranspiration in land-atmosphere interactions. The method could overcome the spatial and temporal coverage limitations of in situ techniques and the inaccuracy of modeling approaches at regional scales. Zhu et al. (2019) predicted global radiative flux and feedbacks from atmospheric and surface variables based on a neural network.

Air pollution is damaging both the earth's environment and human health. Li et al. (2017) used a DNN to fuse satellite observations and station measurements for estimating ground-level PM2.5 levels. Shen et al. (2018) directly estimated the PM2.5 level in Wuhan from the top-of-atmosphere reflectance measured from satellites based on a DNN. Similarly, Tang et al. (2018) used machine learning to estimate the PM10 level from satellite data. They obtained over all accuracy $R^2$=0.77. Weather forecasting is a long-standing challenge in atmospheric science. Han et al. (2017) used machine learning for precipitation nowcasting with radar reflectivity data. This system is superior to traditional expert systems, such as small and easy-to-maintain systems.

Rüttgers et al. (2019) predicted the tracks of typhoons with a GAN based on satellite images. They produced a 6-hour-advance track with an average error of 95.6 km. Jiang et al. (2018) predicted both typhoon track and intensity based on a DNN (Figure 21). Flow-dependent typhoon-induced sea surface temperature cooling was estimated by a DNN and used for improving typhoon predictions.

### 5.5 Remote sensing

Remote sensing uses sensors in satellites or aerial crafts to image geophysical parameters. Remote sensing imagery mainly includes hyperspectral images, SAR images, and optical images. Images obtained by hyperspectral sensors have rich spectral information, such that different land cover categories can potentially be precisely differentiated. In recent years, numerous works have explored deep learning methods for hyperspectral image classification (Li et al. 2019). Chen et al. (2016) used a 3-D CNN to extract the effective features of hyperspectral imagery by considering the spectral-spatial structure simultaneously. The extracted features are useful for image classification and target detection. Mou et al. (2017) first proposed using an RNN for hyperspectral image classification. The authors regarded hyperspectral pixels as sequential data to explore the relationships among different spectrum channels.

SAR systems can operate in all-weather and day-and-night conditions to produce high-resolution images. Chen et al. (2016) used a CNN for target classification in SAR images. The proposed method avoided handcrafted features and achieved an accuracy of 99% in a ten-class classification scheme for ground targets. Zhang et al. (2017) proposed a complex-valued CNN for polarimetric SAR image classification. The authors took both the amplitude and phase information of complex SAR imagery into consideration. The layers in the CNN were extended to process complex-valued inputs. Saha et al. (2020) detect changes in buildings in SAR images via unsupervised learning. A CycleGAN was trained to map SAR images to optical images with no requirement of a paired training set. Then, traditional methods were used to detect building changes.

Large-scale and high-resolution satellite optical color imagery can be used for precision agriculture and urban planning. Maggiori et al. (2017) proposed a CNN-based pixelwise classification of large-scale satellite imagery. Inaccurate data were considered by a two-step training approach. First, the CNN was initialized by numerous inaccurate reference data and then refined on a small amount of correctly labeled data. Cheng et al. (2016) proposed a rotation-invariant CNN for object detection in very high-resolution optical remote sensing images. A rotation-invariant layer was introduced by enforcing the training samples before and after rotation to share the same features. Jiang et al. (2019) constructed an edge-enhancement GAN for remote sensing image superresolution. The image contours were extracted to remove the artifacts and noise in superresolution.

More advanced remote sensing techniques, such as LiDAR, have become popular in recent years. Moorthy et al. (2020) proposed a data-driven method for wood and leaf classification from LiDAR point clouds of forests. Wood and leaf classification in forestry and ecology provides a better understanding of radiation transfer between the canopy and atmosphere. Three tree-based machine learning methods are tested, and they achieved a classification accuracy of up to 94.2%.

### 5.6  Space science

The planets in outer space have attracted researchers' attention for a long time. Camporeale et al. (2017) used machine learning to classify solar wind. The classification results were used to obtain the transition probabilities between different solar wind categories for the first time. Ruhunusiri et al. (2018) developed a DNN to infer solar wind proxies at Mars using sheath measurements. Seven solar wind parameters were inferred simultaneously using spacecraft measurements. Montavon et al. (2020) used neural networks to predict the entire evolution (0-4.5 billion years) of the temperature profile of a Mars-like planet. They used a simple MLP with six parameters as inputs: the reference viscosity, activation energy, activation volume of diffusion creep, enrichment factor of heat-producing elements in the crust, initial temperature of the mantle, and the reference time.

Global space parameter estimation and prediction are long-standing tasks in space science. Chu et al. (2017) used a neural network to predict short-term and long-term 3-D dynamic electron densities in the inner magnetosphere. This network can obtain the magnetospheric plasma density at any time and for any location. Chen et al. (2019) reconstructed dynamic total electron content maps with a regularized GAN. Some existing maps were used as references to interpolate missing values in some regions, such as the oceans. The topside electron temperature is an important parameter of the ionosphere; however, its measurement is limited by the number of incoherent scatter radar stations. Hu et al. (2020) used a DNN to estimate the relationship between electron temperature and electron density in small regions. Therefore, the global electron density is easily measured and used to predict the global electron temperature. Gowtam et al. (2019) modeled a global 3-D ionosphere based on a neural network. Nearly two decades of measurements from the ground and satellites were used to train the neural network. They successfully predicted large-scale ionospheric phenomena, such as annual anomalies.

An aurora is an astronomical phenomenon commonly observed in polar areas. Auroras are caused by disturbances in the magnetosphere caused by solar wind. Auroral classification is important for polar and solar wind research. Clausen and Nickisch (2018) proposed classifying auroral images with a DNN (Figure 22). The authors used images manually labeled with six classes. Zhong et al. (2020) automatically classified all-sky auroral images with three CNNs. The classification results were used to produce an auroral occurrence distribution. Nichols et al. (2019) analyzed Jupiter's auroral morphology and its response to magnetospheric drivers with machine learning. Yang et al. (2019) used a CycleGAN model to extract key local structures from all-sky auroral images. The unpaired training set consisted of 2508 auroral images, of which only 200 images were annotated.

## 6 Revisiting the relationship between dictionary learning and deep learning

In this section, we discuss the relationship between dictionary learning and two specific deep learning methods, the deep image prior (DIP) (Lempitsky et al. 2018) and AE methods. The dictionary learning model in equation (1) is written in a specific form with a sparse constraint on the coefficient:

$$E(\mathbf{W}, \mathbf{C}) = \left\| \mathbf{W}^\mathrm{T} \mathbf{C} - \mathbf{X} \right\|_\mathrm{F} + \alpha \left\| \mathbf{C} \right\|_1$$

where $\alpha$ is a weight parameter. Lempitsky, Vedaldi et al. (2018) proposed DIP, which uses a DNN with random inputs for regularization compared to traditional regularization. DIP uses the target data as the only training samples. Mathematically, DIP replaces the dictionary and the sparse constraint with a single, deep U-Net-based generator.

$$E(\mathbf{\Theta}) = \left\| \underbrace{\text{Generator}(\mathbf{v};\mathbf{\Theta})}_{\mathbf{W}^{\mathrm{T}} \text{ and sparsity}} - \mathbf{X} \right\|_{\mathrm{F}}$$

After obtaining an optimized $\hat{\mathbf{\Theta}}$, another round of forwarding generator propagation will produce a regularized result. In the AE, sparsity is achieved with an encoder that outputs a low-dimensional vector. A decoder corresponds to $\mathbf{W}^{\mathrm{T}}$ in dictionary learning,

$$E(\mathbf{\Theta}_{\mathrm{E}}, \mathbf{\Theta}_{\mathrm{D}}) = \sum_i \left\| \underbrace{\text{Decoder}}_{\mathbf{W}^{\mathrm{T}}}\left( \underbrace{\text{Encoder}(\mathbf{X}_i;\mathbf{\Theta}_{\mathrm{E}})}_{\mathbf{C} \text{ with sparsity}};\mathbf{\Theta}_{\mathrm{D}} \right) - \mathbf{X}_i \right\|_2^2$$

where $\mathbf{\Theta}_{\mathrm{E}}$ and $\mathbf{\Theta}_{\mathrm{D}}$ are the parameters of the encoder and decoder, respectively. Figure 23 shows diagrams of the use of dictionary learning, DIP, and an AE for geophysical signal processing.

## 7 A deep learning tutorial for beginners

### 7.1 A coding example of a DnCNN

The implementation of deep learning algorithms in geophysical data processing is quite simple based on existing frameworks, such as Caffe, Pytorch, Keras, and TensorFlow. Here, we provide an example of how to use Python and Keras to construct a DnCNN for seismic denoising. The code requires 12 lines for dataset loading, model construction, training, and testing. The dataset is preconstructed and includes a clean subset and a noisy subset; the overall dataset includes 12800 samples with size $64 \times 64$ (available at https://bit.ly/33SyXPO).

```
1. import h5py
2. from tensorflow.keras.layers import  Input,Conv2D,BatchNormalization,ReLU,Subtract
3. from tensorflow.keras.models import Model
4. ftrain = h5py.File('noise_dataset.h5','r')
5. X, Y = ftrain['/X'][()] , ftrain['/Y'][()]
6. input = Input(shape=(None,None,1))
7. x = Conv2D(64, 3, padding='same',activation='relu')(input)
8. for i in range(15):
```

```
9.      x = Conv2D(64, 3, padding='same',use_bias = False)(x)
10.     x = ReLU()(BatchNormalization(axis=3, momentum=0.0,epsilon=0.0001)(x))
11. x = Conv2D(1, 3, padding='same',use_bias = False)(x)
12. model = Model(inputs=input, outputs=Subtract()([input, x]))
13. model.compile(optimizer="rmsprop", loss="mean_squared_error")
14. model.fit(X[:-1000], Y[:-1000], batch_size=32, epochs=50, shuffle=True)
15. Y_ = model.predict(X[-1000:])
```

Any appropriate plotting tool can be used for data visualization. The training takes less than one hour on an NVidia 2080ti graphics processing unit. For further implementations, we suggest some public repositories of dictionary learning and deep learning information for interested readers, as listed in Table 2.

### 7.2   Tips for beginners

We introduce some practical tips for beginners who want to explore deep learning in geophysics from the perspective of the three most critical steps in deep learning: data generation, network construction, and training.

#### 7.2.1 Data generation

As noted by Poulton 2002, "training a feed-forward neural network is approximately 10% of the effort involved in an application; deciding on the input and output data coding and creating good training and testing sets is 90% of the work". In deep learning, we advise that the percentages of the effort for network construction and dataset preparation should be approximately 40% and 60%. First, most deep learning approaches use an original data set as the input, thus reducing coding decision efforts. Second, a wider variety of network architectures and parameters can be used in deep learning compared to those in traditional neural networks. Overall, constructing a proper training set plays a more prominent role in deep learning.

Synthetic datasets can be effectively used in deep learning, which is advantageous since realistic datasets with labels are difficult to obtain. First, to assess the availability of deep learning in a specific geophysical application, using synthetic datasets is the most convenient method. Second, if a satisfactory result is obtained with synthetic datasets, realistic datasets can be used for network analysis via transfer learning with a few annotated realistic datasets. Third, if the synthetic datasets are sufficiently complicated, i.e., if the most important factors are considered when generating the datasets, the trained network may be able to process realistic datasets directly (Wu, Geng et al. 2020 and Wu, Liang et al. 2019).

A synthetic training set should be diverse. First, we suggest using an existing synthetic dataset with an open license, such as SEG open data, instead of generating a dataset. For specific tasks, such as FWI, a dataset may need to be generated based on a wave equation. Second, a dataset can be modified to increase the degrees of freedom. For example, noise, missing traces, and faults can be added to clean datasets depending on the considered task. Third, data augmentation can be used to expand a training set, such as via rotation, symmetry, scaling, translation, and other processes. The goal is to generate extremely large synthetic datasets that are as close to realistic datasets as possible.

To generate realistic datasets, we suggest using existing methods to generate labels that should then be checked by a human. For example, in first-arrival picking, an automatic picking algorithm is used to preprocess the datasets, and the results are then provided to an expert who identifies the outliers. Such a procedure requires a human to check very notation. We also suggest using activate learning (Yoo and Kweon 2019) to provide a semiautomated labeling procedure. First, all datasets with machine annotation are used to train a DNN, and the samples with high predicted uncertainty are required to be manually annotated.

### 7.2.2 Network construction for different tasks

Beginners are suggested to use a DnCNN or U-Net for testing. DnCNNs are available for most tasks in which the input and output share the same domain, such as denoising, interpolation, and attribute analysis. The input size of a DnCNN can vary since there are no pooling layers involved. However, each output data point is determined by a local field from the input rather than from the entire input set. Additionally, U-Net contains pooling layers, and all input points are used to determine an output point. U-Nets are available for tasks even when the inputs and outputs are in different domains, such as in FWI. However, the input size of U-Net is fixed once trained.

Combining a CAE and K-means is suggested for unsupervised clustering tasks, such as attribute classification. We do not suggest CycleGAN for geophysical tasks since the training process is extremely time consuming and the results are not stable. An RNN provides a high-performance framework for time-dependent tasks, such as forward wave modeling and FWI. RNNs are also used for regression and classification tasks involving temporal or spatial sequential datasets, such as in the denoising of a single trace.

To adjust the hyperparameters of a DNN and optimization algorithms, we suggest using an autoML toolbox, such as Autokeras, instead of manually adjusting the values. The basic objective is to search for the best parameter combination within a given sampling range. Such a search is exceptionally time consuming, and a random search strategy may accelerate the tuning process. Moreover, for most applications, the default architecture gives reasonable results.

### 7.2.3 Training, validation, and testing

The available dataset should be split into three subsets: one training set, one validation set, and one test set to optimize the network parameters. The proportions are suggested as 60%, 20%, and 20% based on experience. In a classification task, we suggest using one-hot coding in training. The validation set is used to test the network during training. Then, the model with the best validation accuracy is selected rather than the final trained model. If the validation accuracy does not improve during training, an early stopping strategy is suggested to avoid wasted time. Network hyperparameters should be tuned according to the validation accuracy. The validation set is used to guide training, and the test set is used to test the model based on unseen datasets; however, this set should not be used for hyperparameter tuning.

Two commonly seen issues during training are as follows: the validation loss is less than the training loss, and the loss is not a number. Intuitively, the training loss should be less than the validation loss since the model is trained with a training dataset. Some potential reasons for this issue are as follows: 1. regularization occurs during training but is ignored during validation, such as in the dropout layer; 2. the training loss is obtained by averaging the loss of each batch during an iteration, and the validation loss is obtained based on the loss after one iteration; and 3. the validation set may be less complicated than the training set. The potential reasons for NaN loss are as follows: 1. the learning rate is too high; 2. in an RNN, one should clip the gradient to avoid gradient explosion and 3. zero is used as a divisor, negative values are used in logarithm, or an exponent is assigned too large of a value.

## 8    Future directions for deep learning in geophysics

Deep learning, as an efficient artificial intelligence technique, is expected to discover geophysical concepts and inherit expert knowledge through machine-assisted mathematical algorithms. Despite the success of neural networks, their use as a tool for practical geophysics is still in its infancy. The main problems include a shortage of training samples, low signal-to-noise ratios, and strong nonlinearity. Among these issues, the critical challenge is the lack of training samples in geophysical applications compared to those in other industries. Several advanced deep learning methods have been proposed related to this challenge, such as semisupervised and unsupervised learning, transfer learning, multimodal deep learning, federated learning, and active learning. We suggest that a focused be placed on the subjects below for future research in the coming decade.

### 8.1   Semisupervised and unsupervised learning

In practical geophysical applications, obtaining labels for a large dataset is time consuming and can even be infeasible. Therefore, semisupervised or unsupervised learning is required to limit the dependence on labels. Dunham et al. (2019) focused on the application of semisupervised learning in a situation in which the available labels were scarce. A self-training-based label propagation method was proposed, and it outperformed supervised learning methods in which unlabeled samples were neglected. Semisupervised learning takes advantage of both labeled and unlabeled datasets. The combination of AE and K-means is an efficient unsupervised learning method (He, Cao et al. 2018 and Qian, Yin et al. 2018). An autoencoder is used to learn low-dimensional latent features in an unsupervised way, and then K-means is used to cluster the latent features.

## 8.2 Transfer learning

Usually, we must train one DNN for a specific dataset and a specific task. For example, a DNN may effectively process land data but not marine data, or a DNN may be effective in fault detection but not in facies classification. To increase the reusability of a trained network for different datasets or different tasks, transfer learning (Donahue, Jia et al. 2014) is suggested.

In transfer learning with different datasets, the optimized parameters for one dataset can be used as initialization values for learning a new network with another dataset; this process is called fine tuning. Fine tuning is typically much faster and easier than training a network with randomly initialized weights from scratch. In transfer learning involving different tasks, we assume that the extracted features should be the same in different tasks. Therefore, the first layers in a model trained for one task are copied to the new model for another task to reduce the training time. Another benefit of transfer learning is that with a small number of training samples, we can promptly transfer the learned features to a new task or a new dataset. Diagrams of these two transfer learning methods are shown in Figure 24. Further topics in transfer learning include the relationship between the transferability of features (Yosinski et al. 2014) and the distance between different tasks and different data sets (Oquab et al. 2014).

## 8.3 Combination of data-driven and model-driven methods

To combine geophysical mechanics and deep learning, can we combine model-driven and data-driven approaches? Intuitively, such a combination will produce a more precise result than model-driven methods and a more reliable result than data-driven methods. In addition, with an additional physical constraint on deep learning methods, fewer training samples are required to obtain a more generalized prediction than those of traditional methods. Zhang et al. (2017) proposed learning a denoising prior with a DNN and replacing the denoiser in the iteration optimization algorithm, such that different tasks use the same denoiser but different models. Raissi et al. (2019) proposed a physical informed neural network (PINN) that combines training data and physical equation constraints for training. Taking wave modeling as an example, the wavefield was represented with a DNN, $u(x,t) = F(x,t; \mathbf{\Theta})$, such that the acoustic wave equation was:

$$u_{tt} = c^2 \Delta u \xrightarrow{\;u(x,t)=F(x,t;\boldsymbol{\Theta})\;} F_{tt}\left(x,t;\boldsymbol{\Theta}\right) = c^2 \Delta F\left(x,t;\boldsymbol{\Theta}\right)$$

The above equation can serve as a constraint while training the DNN. Tartakovsky et al. (2020) used PINN to learn parameters and constitutive relationships in subsurface flow problems. Another discussed deep learning technique, DIP, can be applied in different tasks with physical models. Similar to the idea of DIP, Wu and McMechan (2019) showed that a DNN generator can be added to an FWI framework. First, a U-Net-based generator $F(\mathbf{v};\boldsymbol{\Theta})$ with random input $\mathbf{v}$ was used to approximate a velocity model $\mathbf{m}$ with high accuracy. Then, $\mathbf{m} = F(\mathbf{v};\boldsymbol{\Theta})$ was inserted into the FWI objective function:

$$\mathrm{E}_{\mathrm{FWI}}(\boldsymbol{\Theta}) = \frac{1}{2}\left\| P(F(\mathbf{v};\boldsymbol{\Theta})) - \mathbf{d}_r \right\|_2^2$$

where $\mathbf{d}_r$ is the seismic record and $P$ is the forward wavefield propagator. The gradient of $\mathrm{E}_{\mathrm{FWI}}$ with respect to network parameters $\boldsymbol{\Theta}$ is calculated with the chain rule. U-Net is only used for regularizing the velocity model. After training, one forward propagation of the network will produce a regularized result. Data-driven and model-driven methods are not independent; data-driven methods are also used for discovering physical concepts (Iten et al. 2020).

### 8.4 Multimodal deep learning

To improve the resolution of inversion, the joint inversion of data from different sources has been a popular topic in recent years (Garofalo et al. 2015). One of the advantages of DNNs is that they can fuse information from multiple inputs. In multimodal deep learning (Ngiam et al. 2011, Ramachandram and Taylor 2017), inputs are from different sources, such as seismic data and gravity data. Collecting data from different sources can help relieve the bottleneck of a limited number of training samples. In addition, using multimodal datasets can increase the accuracy and reliability of deep learning methods. Feng et al. (2020) used data integration to forcast streamflow. 23 variables were used intergrated, such as precipitation, solar radiation, and temperature. Figure 25 shows an illustration of multimodal deep learning.

### 8.5 Federated learning

To provide a practical training set in deep learning for geophysical applications, collecting available datasets from different institutes or corporations might be a possible solution. However, data transfer via the internet is time consuming and expensive for large-scale geophysical datasets. In addition, most datasets are protected and cannot be shared. Federated learning was first proposed by Google (Mcmahan et al. 2017, Li et al. 2020) to train a DNN with user data from millions of cellphones without privacy or security issues. The encrypted gradients from different clients are assembled in a central server, thus avoiding data transfer. The server updates the model and distributes information to all clients (Figure 26). In a simple federated learning setting, the clients and the server share the same network architecture. We give a possible example of federated learning in geophysics based on the concept that some corporations do not share the annotations of first arrivals; however, they can benefit from federated learning by training a DNN together for first arrival picking.

### 8.6 Uncertainty estimation

One of the remaining questions associated with applying deep learning in geophysics is related to whether the results of deep learning-based model-driven methods with a solid theoretical foundation can be trusted. One trial in drilling may cost millions of dollars. What if a neural network can report high confidence in a prediction? Deep learning with uncertainty analysis was proposed to assess reliability, such as through Markov chain Monte Carlo (MCMC) (de Figueiredo et al. 2019), variational inference (Subedar et al. 2019), and Monte Carlo dropout (Gal and Ghahramani 2016) methods. For example, in Monte Carlo dropout, dropout layers are added to each original layer to simulate a Bernoulli distribution. With multiple realizations of dropout, the results are collected, and the variance is computed as the uncertainty.

Grana et al. (2020) assessed the classification accuracy and uncertainty of RNN and MCMC methods. The RNN method yielded higher accuracy but relatively high uncertainty. The MCMC method provided similar accuracy and was robust to uncertainty through the use of prior spatial correlation models. In the RNN, the uncertainty was obtained through multiple runs of the same procedure with different training subsets, but the results were similar in each case. Maiti and Tiwari (2010) used a Bayesian neural network (BNN) to predict the boundaries of lithofacies. BNNs provide low uncertainty compared to traditional deep learning methods. Cao et al. (2020) proposed a sequence of fast seismic acquisitions for dispersion curve extraction and inversion for 3-D seismic models with uncertainty estimates using pretrained mixture density networks.

### 8.7    Active learning

To train a high-precision model using a small amount of labeled data, active learning is proposed to imitate the self-learning ability of human beings (Yoo and Kweon 2019). An active learning model selects the most useful data based on a sampling strategy for manual annotation and adds this data to the training set; then, the updated dataset is used for the next round of training (Figure 27). One of the sampling strategies is based on the uncertainty principle, i.e., the samples with high uncertainty are selected. Taking fault detection as an example, if a trained network is not sure whether a fault exists at a given location, we can annotate the fault manually and add the sample to the training set.

## 9    Summary

Data-driven methods, especially deep learning methods, have created both opportunities and challenges in geophysical fields. Pioneer researchers have provided a basis for deep learning in geophysics with promising results; more advanced deep learning technologies and more practical problems must now be explored. To close this paper, we summarize a roadmap for applying deep learning in different geophysical tasks based on a three-level approach.

- Traditional methods are time consuming and require intensive human labor and expert knowledge, such as in first-arrival selection and velocity selection in explorational geophysics.

- Traditional methods have difficulties and bottlenecks. For example, geophysical inversion requires good initial values and high accuracy modeling and suffers from local minimization.

- Traditional methods cannot handle some cases, such as multimodal data fusion and inversion.

With the development of new artificial intelligence models beyond deep learning and advances in research into the infinite possibilities of applying deep learning in geophysics, we can expect intelligent and automatic discovering of unknown geophysical principles soon.

## Glossary

AE: Autoencoder; an ANN with the same inputs and outputs.

AI: Artificial Intelligence; Machines are taught to think as humans.

ANN: Artificial neural network; a computing system inspired by biological neural networks that constitute animal brains.

Aurora: A natural light display in the earth's sky; disturbances in the magnetosphere caused by solar wind.

BNN: Bayesian neural network; the network parameters are random variables instead of regular variables.

CAE: Convolutional autoencoder; an AE with shared weights.

CNN: Convolutional neural network; a DNN with shared weights.

Compressive sensing: A sampling technique for reconstructing a signal with a sample rate lower than the requirement of Shannon sampling theory. The mathematical principle of compressive sensing is that an underdetermined linear system can be solved with a sparse prior on the signal.

DDTF: Data-driven tight frame; A dictionary learning method using a tight frame constraint for the dictionary.

Deblending: In seismic exploration, several explosion sources are shot very close in time to improve efficiency. Then, the seismic waves from different sources are blended. The recorded dataset first needs to be deblended before further processing.

Deduction: The principles used for the prediction of unknown facts.

Dictionary: A set of vectors used to represent signals as a linear combination.

DIP: Deep image prior; the architecture of a DNN is used as a prior constraint for an image.

DL: Deep learning; a machine learning technology based on a deep neural network.

DnCNN: Denoised convolutional neural network.

DNN: Deep neural network; an ANN with many layers between the input and output layers.

DS: Double sparsity; the data are represented with a sparse coefficient matrix multiplied by an adaptive dictionary. The adaptive dictionary is represented by a sparse coefficient matrix multiplied by a fixed dictionary.

EEW: Earthquake early warning; earthquake alerts are sent to people seconds to tens of seconds in advance of when shaking waves are expected to arrive. EEW systems make use of the speed difference between propagation of the primary and secondary waves.

Event: In exploration geophysics, a seismic event means reflected waves with the same phase. In seismology, an event means a happened earthquake.

Facies: A seismic facies unit is a mapped, three-dimensional seismic unit composed of groups of reflections whose parameters differ from adjacent facies units.

Fault: a discontinuity in a volume of rock across which there has been significant displacement as a result of rock-mass movement.

FCN: Fully convolutional network; an FCN is a network that contains no fully connected layers. Fully connected layers do not share weights.

FWI: Full waveform inversion; full waveform information is used to obtain subsurface parameters. FWI is achieved based on the wave equation and inversion theory.

GAN: Generative adversarial network; GANs are used to generate fake images. A GAN contains a generative network and a discriminative network. The generative network tries to produce a nearly real image. The discriminative network tries to distinguish whether the input image is real or generated. Therefore, such a game will eventually allow the generative network to produce fake images that the discriminative network cannot distinguish from real images.

Graphics processing unit (GPU): A parallel computing device. GPUs are widely used for training neural works in deep learning.

HadCRUT4: Temperature records from Hadley Centre (sea surface temperature) and the Climatic Research Unit (land surface air temperature).

Induction: Principles are inferred from observations.

Ionosphere: The ionized part of the earth's upper atmosphere.

K-means: A classical clustering algorithm, where K is the number of clusters.

K-SVD: A dictionary learning method using SVD for dictionary updating.

Kullback-Leibler divergence: A measurement of the distance between two probability distributions

LiDAR: A measurement device using laser light to produce high-resolution images.

LSTM: long short-term memory; LSTM considers how much historical information is forgotten or remembered with adaptive switches.

Magnetosphere: Range of the magnetic field surrounding an astronomical object where charged particles are affected.

$M_L$: Earthquake local magnitude; a method for measuring earthquake scale.

MLP: Multilayer perceptron; A feedforward ANN in which only adjacent layers are connected.

MOD: Method of optimal directions; a dictionary learning method using orthogonal matching pursuit for sparse coding.

Patch: In dictionary learning, an image is divided into many patches (blocks) that are the same size as the atoms in a dictionary.

PINN: Physical informed neural network; A physical equation is used to constrain the neural network.

PM: Particulate matter. PM10 are coarse particles with a diameter of 10 micrometers or less; PM2.5 are fine particles with a diameter of 2.5 micrometers or less.

$R^2$: Coefficient of determination; A statistical parameter range from zero to one indicates how strong the linear relationship is between two variables.

ResNet: Residual neural network; ResNets contain skip connections to jump over some layers. The output of a residual block is the residual between the input and the direct output.

RNN: Recurrent neural network; in time-sequenced data processing applications, RNNs use the output of a network as the input of the subsequent process to consider the historical context.

SAR: Synthetic aperture radar; the motion of a radar antenna over a target is treated as an antenna with a large aperture. The larger the aperture is, the higher the image resolution will be.

Sequencer: A machine learning-based ordering algorithm in which similar waveforms are close to each other.

Solar wind: A stream of charged particles released from the upper atmosphere of the Sun.

Sparse coding: Input data are represented in the form of a linear combination of a dictionary where the coefficients are sparse.

Sparsity: The number of nonzero values in a vector.

SVD: Singular value decomposition; a matrix factorization method. **A=USV**, where **U** and **V** are two orthogonal matrices, **S** is a diagonal matrix whose elements are the singular values of **A**. SVD is used for dimension reduction by removing the smaller singular values. SVD is also used for recommendation systems and natural language processing.

Tight frame: A frame provides a redundant, stable way of representing a signal, similar to dictionary. A tight frame is a frame with the perfect reconstruction property; i.e., $\mathbf{W}^T\mathbf{W}=\mathbf{I}$.

Tomography: Inversion of the subsurface velocity based on travel time information.

U-Net: U-shaped network; U-Nets have U-shaped structures and skip connections. The skip connections bring low-level features to high levels.

Velocity analysis: Analysis of a velocity distribution along the depth based on signal semblance.

Wave equation: A partial differential equation that controls wave propagation.

WST: Wavelet scattering transform; a transform involves a cascade of wavelet transforms, a module operator, and an averaging operator.

## Acknowledgments

The work was supported in part by the National Key Research and Development Program of China under grant nos. 2017YFB0202902 and 2018YFC1503705 and NSFC under grant nos. 41625017 and 41804102. We thank Society of Exploration Geophysicists, Nature Research, and American Association for the Advancement of Science for allowing us to reuse the original figures from their journals.

## Data Availability Statement

Data were not used, nor created for this research.

## References

Abma, R. and N. Kabir (2006). 3D interpolation of irregular data with a POCS algorithm. Geophysics. **71**(6): 91-97.

Aharon, M., M. Elad and A. Bruckstein (2006). K-SVD: An algorithm for designing overcomplete dictionaries for sparse representation. IEEE Transactions on Signal Processing. **54**(11): 4311-4322.

Araya-Polo, M., J. Jennings, A. Adler and T. Dahlke (2018). Deep-learning tomography. The Leading Edge. **37**(1): 58-66.

Beckouche, S. and J. Ma (2014). Simultaneous dictionary learning and denoising for seismic data. Geophysics. **79**(3): A27-A31.

Cai, J., H. Ji, Z. Shen and G. Ye (2014). Data-driven tight frame construction and image denoising. Applied and Computational Harmonic Analysis. **37**(1): 89-105.

Cao, R., S. Earp, S. A. L. de Ridder, A. Curtis and E. Galetti (2020). Near-real-time near-surface 3D seismic velocity and uncertainty models by wavefield gradiometry and neural network inversion of ambient seismic noise. Geophysics. **85**(1): KS13-KS27.

Chen, H., R. Guo, J. Liu, Y. Wang and R. Lin (2020). Magnetotelluric data denoising with recurrent neural network. SEG 2019 Workshop: Mathematical Geophysics: Traditional vs Learning.

Clausen, L. B. N. and H. Nickisch (2018). Automatic classification of auroral images from the oslo auroral themis (OATH) data set using machine learning. Journal of Geophysical Research-Space Physics. **123**(7): 5640-5647.

Cortes, C. and V. Vapnik (1995). Support-vector networks. Machine learning. **20**(3): 273-297.

Creswell, A., T. White, V. Dumoulin, K. Arulkumaran, B. Sengupta and A. A. Bharath (2018). Generative adversarial networks: An overview. IEEE Signal Processing Magazine. **35**(1): 53-65.

Das, V., A. Pollack, U. Wollner and T. Mukerji (2019). Convolutional neural network for seismic impedance inversion. Geophysics. **84**(6): R869-R880.

de Figueiredo, L. P., D. Grana, M. Roisenberg and B. B. Rodrigues (2019). Gaussian mixture markov chain Monte Carlo method for linear seismic inversion. Geophysics. **84**(3): R463-R476.

De Lima, R. P. and K. J. Marfurt (2018). Principal component analysis and K-means analysis of airborne gamma-ray spectrometry surveys. SEG Technical Program Expanded Abstracts: 2277-2281.

Donahue, J., Y. Jia, O. Vinyals, J. Hoffman and T. Darrell (2014). Decaf: A deep convolutional activation feature for generic visual recognition. International Conference on Machine Learning: 647-655.

Dong, C., C. C. Loy, K. He and X. Tang (2014). Learning a deep convolutional network for image super-resolution. European Conference on Computer Vision: 184-199.

Dragomiretskiy, K. and D. Zosso (2014). Variational mode decomposition. IEEE Transactions on Signal Processing. **62**(3): 531-544.

Duan, Y., X. Zheng, L. Hu and L. Sun (2019). Seismic facies analysis based on deep convolutional embedded clustering. Geophysics. **84**(6): IM87-IM97.

Dunham, M. W., A. Malcolm and J. Kim Welford (2019). Improved well-log classification using semisupervised label propagation and self-training, with comparisons to popular supervised algorithms. Geophysics. **85**(1): O1-O15.

Engan, K., S. O. Aase and J. H. Husoy (2002). Method of optimal directions for frame design. 1999 IEEE International Conference on Acoustics, Speech, and Signal Processing. Proceedings. ICASSP99 (Cat. No.99CH36258).

Evgeniou, T., M. Pontil and T. Poggio (2000). Regularization networks and support vector machines. Advances in Computational Mathematics. **13**(1): 1-50.

Fabien-Ouellet, G. and R. Sarkar (2019). Seismic velocity estimation: A deep recurrent neural-network approach. Geophysics. **85**(1): U21-U29.

Feng, D., K. Fang and C. Shen (2020). Enhancing streamflow forecast and extracting insights using long-short term memory networks with data integration at continental scales. Water Resources Research.

Gal, Y. and Z. Ghahramani (2016). Dropout as a Bayesian approximation: Representing model uncertainty in deep learning. International Conference on Machine Learning.

Galvis, I. S., Y. Villa, C. Duarte, D. A. Sierra and W. Agudelo (2017). Seismic attribute selection and clustering to detect and classify surface waves in multicomponent seismic data by using K-means algorithm. Geophysics. **36**(3): 239-248.

Gao, Z., Z. Pan, J. Gao and Z. Xu (2019). Building long-wavelength velocity for salt structure using stochastic full waveform inversion with deep autoencoder based model reduction. SEG Technical Program Expanded Abstracts: 1680-1684.

Garofalo, F., G. Sauvin, L. V. Socco and I. Lecomte (2015). Joint inversion of seismic and electric data applied to 2D media. Geophysics. **80**(4): EN93-EN104.

Goodfellow, I., J. Pouget-Abadie, M. Mirza, B. Xu, D. Warde-Farley, S. Ozair, A. Courville and Y. Bengio (2014). Generative adversarial nets. Neural Information Processing Systems: 2672-2680.

Grana, D., L. Azevedo and M. Liu (2020). A comparison of deep machine learning and Monte Carlo methods for facies classification from seismic data. Geophysics. **85**(4): WA41-WA52.

Guillon, H., C. F. Byrne, B. A. Lane, S. S. Solis and G. B. Pasternack (2020). Machine learning predicts reach-scale channel types from coarse-scale geospatial data in a large river basin. Water Resources Research. **56**(3).

Hartigan, J. A. and M. A. Wong (1979). A k-means clustering algorithm. Journal of The Royal Statistical Society Series C-applied Statistics. **28**(1): 100-108.

He, K., X. Zhang, S. Ren and J. Sun (2016). Deep residual learning for image recognition. IEEE Conference on Computer Vision and Pattern Recognition: 770-778.

He, Y., J. Cao, Y. Lu, Y. Gan and S. Lv (2018). Shale seismic facies recognition technology based on sparse autoencoder. International Geophysical Conference.

Helmy, T., A. Fatai and K. Faisal (2010). Hybrid computational models for the characterization of oil and gas reservoirs. Expert Systems with Applications. **37**(7): 5353-5363.

Herrmann, F. J. and G. Hennenfent (2008). Non-parametric seismic data recovery with curvelet frames. Geophysical Journal International. **173**(1): 233-248.

Hochreiter, S. and J. Schmidhuber (1997). Long short-term memory. Neural Comput. **9**(8): 1735-1780.

Hu, L., X. Zheng, Y. Duan, X. Yan, Y. Hu and X. Zhang (2019). First-arrival picking with a U-net convolutional network. Geophysics. **84**(6): U45-U57.

Huang, K., J. You, K. Chen, H. Lai and A. Don (2006). Neural network for parameters determination and seismic pattern detection. SEG Technical Program Expanded Abstracts: 2285-2289.

Iten, R., T. Metger, H. Wilming, L. Del Rio and R. Renner (2020). Discovering physical concepts with neural networks. Phys Rev Lett. **124**(1): 010508.

Jia, Y. and J. Ma (2017). What can machine learning do for seismic data processing? An interpolation application. Geophysics. **82**(3): V163-V177.

Jiang, G.-q., J. Xu and J. Wei (2018). A deep learning algorithm of neural network for the parameterization of typhoon-ocean feedback in typhoon forecast models. Geophysical Research Letters. **45**(8): 3706-3716.

Kadow, C., D. M. Hall and U. Ulbrich (2020). Artificial intelligence reconstructs missing climate information. Nature Geoscience. **13**(6): 408-413.

Kim, D., V. Lekic, B. Menard, D. Baron and M. Taghizadeh-Popp (2020). Sequencing seismograms: A panoptic view of scattering in the core-mantle boundary region. Science. **368**(6496): 1223-1228.

Krizhevsky, A., I. Sutskever and G. E. Hinton (2017). Imagenet classification with deep convolutional neural networks. Communications of the Acm. **60**(6): 84-90.

LeCun, Y., B. Boser, J. S. Denker, D. Henderson and L. D. Jackel (1997). Handwritten digit recognition with a back-propagation network. Advances in Neural Information Processing Systems. **2**(2): 396-404.

Lempitsky, V., A. Vedaldi and D. Ulyanov (2018). Deep image prior. IEEE Conference on Computer Vision and Pattern Recognition: 9446-9454.

Li, L., Y. Lin, X. Zhang, H. Liang, W. Xiong and S. Zhan (2019). Convolutional recurrent neural networks based waveform classification in seismic facies analysis. SEG Technical Program Expanded Abstracts: 2599-2603.

Li, T., A. K. Sahu, A. Talwalkar and V. Smith (2020). Federated learning: Challenges, methods, and future directions. IEEE Signal Processing Magazine. **37**(3): 50-60.

Liang, J., J. Ma and X. Zhang (2014). Seismic data restoration via data-driven tight frame. Geophysics. **79**(3): V65-V74.

Lim, J. S. (2005). Reservoir properties determination using fuzzy logic and neural networks from well data in offshore korea. Journal of Petroleum Science and Engineering. **49**(3-4): 182-192.

Lipari, V., F. Picetti, P. Bestagini and S. Tubaro (2018). A generative adversarial network for seismic imaging applications. SEG Technical Program Expanded Abstracts: 2231-2235.

Liu, L. and J. Ma (2019). Structured graph dictionary learning and application on the seismic denoising. IEEE Transactions on Geoscience and Remote Sensing. **57**(4): 1883-1893.

Liu, L., J. Ma and G. Plonka (2018). Sparse graph-regularized dictionary learning for suppressing random seismic noise. Geophysics. **83**(3): V215-V231.

Liu, L., G. Plonka and J. Ma (2017). Seismic data interpolation and denoising by learning a tensor tight frame. Inverse Problems. **33**(10): 105011.

Liu, S. (2020). Multi-parameter full waveform inversions based on recurrent neural networks. Dissertation for the Master Degree in Science, Harbin Institute of Technology.

Maiti, S. and R. K. Tiwari (2010). Neural network modeling and an uncertainty analysis in Bayesian framework: A case study from the KTB borehole site. Journal of Geophysical Research-Solid Earth. **115**: 1-28.

Mandelli, S., F. Borra, V. Lipari, P. Bestagini, A. Sarti and S. Tubaro (2018). Seismic data interpolation through convolutional autoencoder. SEG Technical Program Expanded Abstracts: 4101-4105.

McCann, M. T., K. H. Jin and M. Unser (2017). Convolutional neural networks for inverse problems in imaging: A review. IEEE Signal Processing Magazine. **34**(6): 85-95.

Mcmahan, H. B., E. Moore, D. Ramage, S. Hampson and B. A. Y. Arcas (2017). Communication-efficient learning of deep networks from decentralized data. International Conference on Artificial Intelligence and Statistics.

Nakayama, S., G. Blacquiere and T. Ishiyama (2019). Automated survey design for blended acquisition with irregular spatial sampling via the integration of a metaheuristic and deep learning. Geophysics. **84**(4): P47-P60.

Nazari Siahsar, M. A., S. Gholtashi, A. R. Kahoo, W. Chen and Y. Chen (2017). Data-driven multitask sparse dictionary learning for noise attenuation of 3D seismic data. Geophysics. **82**(6): V385-V396.

Ngiam, J., A. Khosla, M. Kim, J. Nam, H. Lee and A. Y. Ng (2011). Multimodal deep learning. International Conference on Machine Learning.

Oquab, M., L. Bottou, I. Laptev and J. Sivic (2014). Learning and transferring mid-level image representations using convolutional neural networks. IEEE Conference on Computer Vision and Pattern Recognition.

Ovcharenko, O., V. Kazei, M. Kalita, D. Peter and T. Alkhalifah (2019). Deep learning for low-frequency extrapolation from multioffset seismic data. Geophysics. **84**(6): R989-R1001.

Park, M. J. and M. D. Sacchi (2019). Automatic velocity analysis using convolutional neural network and transfer learning. Geophysics. **85**(1): V33-V43.

Payani, A., F. Fekri, G. Alregib, M. Mohandes and M. Deriche (2019). Compression of seismic signals via recurrent neural networks: Lossy and lossless algorithms. SEG Technical Program Expanded Abstracts 2019: 4082-4086.

Poulton, M. M. (2002). Neural networks as an intelligence amplification tool: A review of applications. Geophysics. **67**(3): 979-993.

Qi, J., B. Zhang, B. Lyu and K. Marfurt (2020). Seismic attribute selection for machine-learning-based facies analysis. Geophysics. **85**(2): O17-O35.

Qian, F., M. Yin, X. Liu, Y. Wang, C. Lu and G. Hu (2018). Unsupervised seismic facies analysis via deep convolutional autoencoders. Geophysics. **83**(3): A39-A43.

Raissi, M., P. Perdikaris and G. E. Karniadakis (2019). Physics-informed neural networks: A deep learning framework for solving forward and inverse problems involving nonlinear partial differential equations. Journal of Computational Physics. **378**: 686-707.

Ramachandram, D. and G. W. Taylor (2017). Deep multimodal learning: A survey on recent advances and trends. IEEE Signal Processing Magazine. **34**(6): 96-108.

Ronneberger, O., P. Fischer and T. Brox (2015). U-net: Convolutional networks for biomedical image segmentation. Medical Image Computing and Computer Assisted Intervention: 234-241.

Rubinstein, R., M. Zibulevsky and M. Elad (2010). Double sparsity: Learning sparse dictionaries for sparse signal approximation. IEEE Transactions on Signal Processing. **58**(3): 1553-1564.

Si, X., Y. Yuan, F. Ping, Y. Zheng and L. Feng (2020). Ground roll attenuation based on conditional and cycle generative adversarial networks. SEG 2019 Workshop: Mathematical Geophysics: Traditional vs Learning.

Siahkoohi, A., M. Louboutin and F. J. Herrmann (2019). The importance of transfer learning in seismic modeling and imaging. Geophysics. **84**(6): A47-A52.

Simonyan, K. and A. Zisserman (2015). Very deep convolutional networks for large-scale image recognition. International Conference on Learning Representations.

Song, C., Z. Liu, Y. Wang, F. Xu, X. Li and G. Hu (2018). Adaptive phase K-means algorithm for waveform classification. Exploration Geophysics. **49**(2): 213-219.

Subedar, M., R. Krishnan, P. L. Meyer, O. Tickoo and J. Huang (2019). Uncertainty-aware audiovisual activity recognition using deep Bayesian variational inference. International Conference on Computer Vision: 6300-6309.

Sun, J., Z. Niu, K. A. Innanen, J. Li and D. O. Trad (2020). A theory-guided deep-learning formulation and optimization of seismic waveform inversion. Geophysics. **85**(2): R87-R99.

Sun, J., S. Slang, T. Elboth, T. Larsen Greiner, S. McDonald and L.-J. Gelius (2020). A convolutional neural network approach to deblending seismic data. Geophysics. **85**(4): WA13-WA26.

Tartakovsky, A. M., C. O. Marrero, P. Perdikaris, G. D. Tartakovsky and D. Barajas-Solano (2020). Physics-informed deep neural networks for learning parameters and constitutive relationships in subsurface flow problems. Water Resources Research. **56**(5).

Waheed, U. B., S. Alzahrani and S. M. Hanafy (2019). Machine learning algorithms for automatic velocity picking: K-means vs. DBSCAN. SEG Technical Program Expanded Abstracts: 5510-5114.

Wang, B., N. Zhang, W. Lu and J. Wang (2019). Deep-learning-based seismic data interpolation: A preliminary result. Geophysics. **84**(1): V11-V20.

Wang, T., Z. Zhang and Y. Li (2019). Earthquakegen: Earthquake generator using generative adversarial networks. SEG Technical Program Expanded Abstracts: 2674-2678.

Wang, W. and J. Ma (2020). Velocity model building in a crosswell acquisition geometry with image-trained artificial neural network. geophysics. **85**(2): U31-U46.

Wang, X. and J. Ma (2019). Adaptive dictionary learning for blind seismic data denoising. IEEE Geoscience and Remote Sensing Letters: 1-5.

Wang, X., B. Wen and J. Ma (2019). Denoising with weak signal preservation by group-sparsity transform learning. Geophysics. **84**(6): V351-V368.

Wang, Y., Q. Ge, W. Lu and X. Yan (2019). Seismic impedance inversion based on cycle-consistent generative adversarial network. SEG Technical Program Expanded Abstracts: 2498-2502.

Wang, Y., B. Wang, N. Tu and J. Geng (2020). Seismic trace interpolation for irregularly spatial sampled data using convolutional autoencoder. Geophysics. **85**(2): V119-V130.

Wu, H., B. Zhang, F. Li and N. Liu (2019). Semiautomatic first-arrival picking of microseismic events by using the pixel-wise convolutional image segmentation method. Geophysics. **84**(3): V143-V155.

Wu, H., B. Zhang, T. Lin, D. Cao and Y. Lou (2019). Semiautomated seismic horizon interpretation using the encoder-decoder convolutional neural network. Geophysics. **84**(6): B403-B417.

Wu, H., B. Zhang, T. Lin, F. Li and N. Liu (2019). White noise attenuation of seismic trace by integrating variational mode decomposition with convolutional neural network. Geophysics. **84**(5): V307-V317.

Wu, X., Z. Geng, Y. Shi, N. Pham, S. Fomel and G. Caumon (2020). Building realistic structure models to train convolutional neural networks for seismic structural interpretation. Geophysics. **85**(4): WA27-WA39.

Wu, X., L. Liang, Y. Shi and S. Fomel (2019). FaultSeg3D: Using synthetic data sets to train an end-to-end convolutional neural network for 3D seismic fault segmentation. Geophysics. **84**(3): IM35-IM45.

Wu, Y. and G. A. McMechan (2019). Parametric convolutional neural network-domain full-waveform inversion. Geophysics. **84**(6): R881-R896.

Yang, F. and J. Ma (2019). Deep-learning inversion: A next-generation seismic velocity model building method. Geophysics. **84**(4): R585-R584.

Yoo, D. and I. S. Kweon (2019). Learning loss for active learning. IEEE Conference on Computer Vision and Pattern Recognition.

Yosinski, J., J. Clune, Y. Bengio and H. Lipson (2014). How transferable are features in deep neural networks. Neural Information Processing Systems.

You, N., Y. E. Li and A. Cheng (2020). Shale anisotropy model building based on deep neural networks. Journal of Geophysical Research: Solid Earth. **125**(2): e2019JB019042.

Yu, S., J. Ma and S. Osher (2016). Monte Carlo data-driven tight frame for seismic data recovery. Geophysics. **81**(4): V327-V340.

Yu, S., J. Ma and W. Wang (2019). Deep learning for denoising. Geophysics. **84**(6): V333-V350.

Yu, S., J. Ma, X. Zhang and M. Sacchi (2015). Interpolation and denoising of high-dimensional seismic data by learning a tight frame. Geophysics. **80**(5): V119-V132.

Zhang, C., C. Frogner, M. Araya-Polo and D. Hohl (2014). Machine-learning based automated fault detection in seismic traces. 76th EAGE Conference and Exhibition 2014. **2014**(1): 1-5.

Zhang, H., W. Wang, X. Wang, W. Chen and Z. Zhao (2019). An implementation of the seismic resolution enhancing network based on GAN. SEG Technical Program Expanded Abstracts 2019: 2478-2482.

Zhang, H., X. Yang and J. Ma (2020). Can learning from natural image denoising be used for seismic data interpolation? Geophysics. **85**(4): WA115-WA136.

Zhang, K., W. Zuo, Y. Chen, D. Meng and L. Zhang (2017). Beyond a Gaussian denoiser: Residual learning of deep CNN for image denoising. IEEE Transactions on Image Processing. **26**(7): 3142-3155.

Zhang, K., W. Zuo, S. Gu and L. Zhang (2017). Learning deep CNN denoiser prior for image restoration. IEEE Conference on Computer Vision and Pattern Recognition: 2808-2817.

Zhang, X., J. Zhang, C. Yuan, S. Liu, Z. Chen and W. Li (2020). Locating induced earthquakes with a network of seismic stations in oklahoma via a deep learning method. Scientific Reports. **10**(1): 1941.

Zhang, Z. and T. Alkhalifah (2019). Regularized elastic full-waveform inversion using deep learning. Geophysics. **84**(5): R741-R751.

Zhu, J., T. Park, P. Isola and A. A. Efros (2017). Unpaired image-to-image translation using cycle-consistent adversarial networks. International Conference on Computer Vision: 2242-2251.

Zu, S., J. Cao, S. Qu and Y. Chen (2020). Iterative deblending for simultaneous source data using the deep neural network. Geophysics. **85**(2): V131-V141.

# Tables

Table 1 Data-driven tasks in Geophysics

| Data-driven Tasks in Geophysics | | |
|---|---|---|
| **Spatial prediction** | Reconstruction | Global climate information based on limited measurements |
| | | All-sky information from limited astronomy observation stations |
| | | Both high resolution and large scale measurement in remote sensing |
| | Inversion | High resolution subsurface structure with surface seismic data in exploration geophysics |
| | | The Earth's structure based on global earthquake measurements |
| **Temporal prediction** | Forward prediciton | Rain fall nowcasting |
| | | Early earthquake warning |
| | | Typhoon track prediction |
| | | Other natural disasters prediction in small time window |
| | Backward prediction | The evolution of the Earth and the Universe in very large time window |
| | | The drift of the continental |
| **Detection** | | Microearthquake detection |
| | | Pond coverage on Arctic sea ice |
| | | Size of icebergs |
| | | Coastal inundation mapping |
| **Classification** | | Large spatial scale remote sensing imagery classification, SAR, Lidar, Optical |
| | | Water stream classification |
| | | Auroal classification |
| | | Solar wind classification |

Table 2 Open sources of dictionary learning and deep learning for geophysical applications

| Open source codes | | |
|---|---|---|
| | | **links** |
| **Dictionary learning** | https://github.com/sevenysw/MathGeo2020/tree/master/DDTF3D | — Denoise <br> — Interpolation |
| **Deep learning** | Denoise —— https://github.com/sevenysw/MathGeo2020/tree/master/Deep_learning/python_segy | |
| | Interpolation —— https://github.com/sevenysw/MathGeo2020/tree/master/Deep_learning/CNN-POCS | |
| | FWI —— https://github.com/sevenysw/MathGeo2020/tree/master/Deep_learning/FCNVMB | |
| | Arrival picking —— https://github.com/DaloroAT/first_break_picking <br> https://github.com/mingzhaochina/ConvNetQuake | |
| | Fault detection —— https://github.com/xinwucwp/faultSeg | |
| | Facies classification —— https://github.com/JesperDramsch/seismic-transfer-learning | |
| | Phase association —— https://github.com/interseismic/PhaseLink | |
| | Integrated Toolkit —— https://github.com/microsoft/seismic-deeplearning | — seismic imaging <br> — interpretation |
| | https://github.com/learnserd/SeismicPro | — Ground-roll attenuation <br> — First-break picking |
| | https://github.com/gazprom-neft/seismiqb | — Horizon segmentations <br> — Interlayers segmentation |
| | https://github.com/maihao14/Lina-Seismic-Playground | — Data denoise <br> — First-break picking |

## Figures

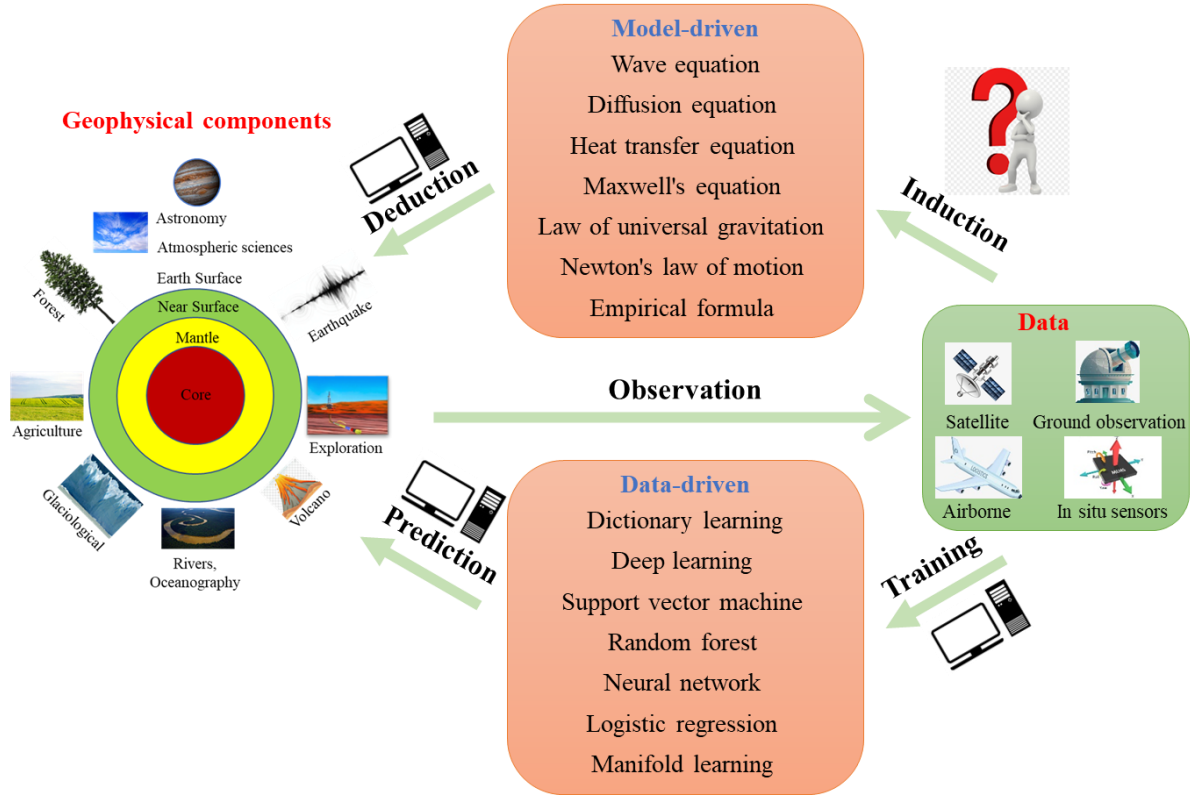

Figure 1 An illustration of model-driven and data-driven methods. On the left are the research topics in geophysics ranging from the Earth's core to the outer space. One the right are the observation means used in model technology. In the middle are examples of model-driven and data-driven methods. In model-driven methods, the principles of geophysical phenomena are induced from a large amount of observed data based on physical causality, then the models are used to deduct the geophyscial phanomena in the future or in the past. In data-driven methods, the computer first induct a regression or classification model without considering physical causality. Then, this model will perform tasks such as prediction, classification, and recognition on incoming datasets.

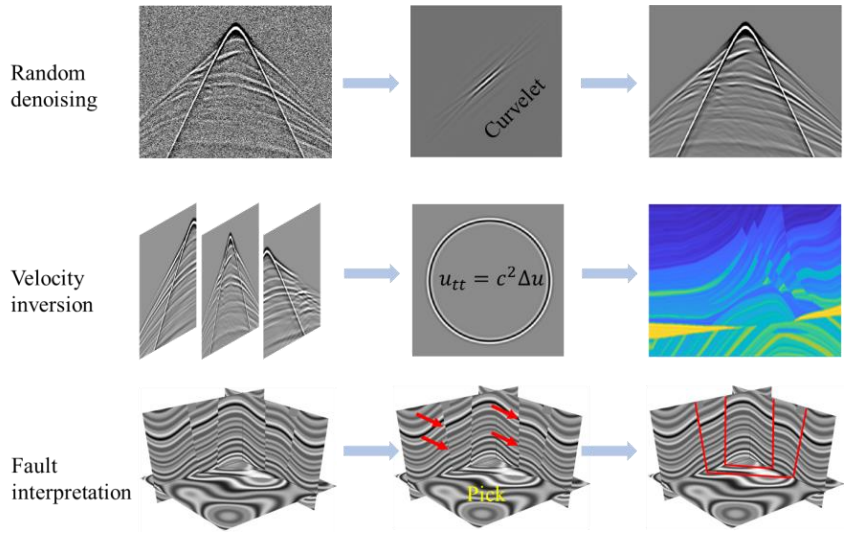

(a) Model-driven methods

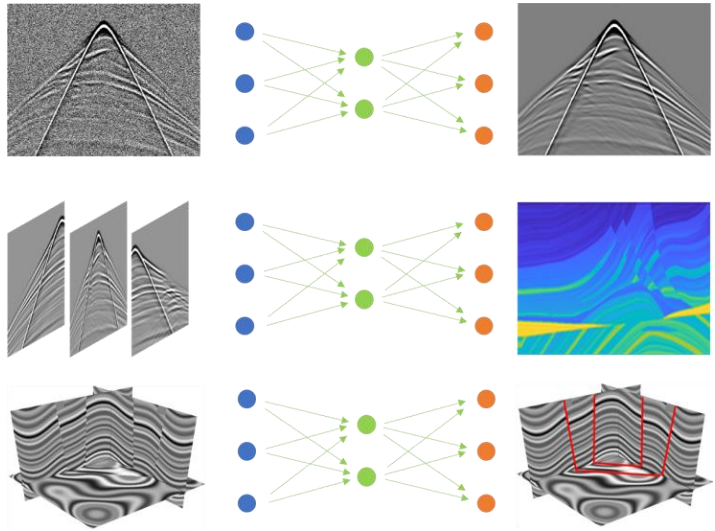

(b) Data-driven methods

Figure 2. Examples of model-driven and data-driven methods in exploration geophysics. (a) Model-driven methods. In random denoising tasks, the curvelet denoising method (Herrmann and Hennenfent 2008) assumes that the signal is sparse under curvelet transform, and a matching method is used for denoising. In velocity inversion tasks, full-waveform inversion based on the wave equation is used for forward and adjoint modeling in the optimization algorithm. In fault interpretation tasks, faults are picked by interpreters. (b) Data-driven methods. The mentioned tasks are treated as regression problems that are optimized with neural networks. Different tasks may require different neural network architectures.

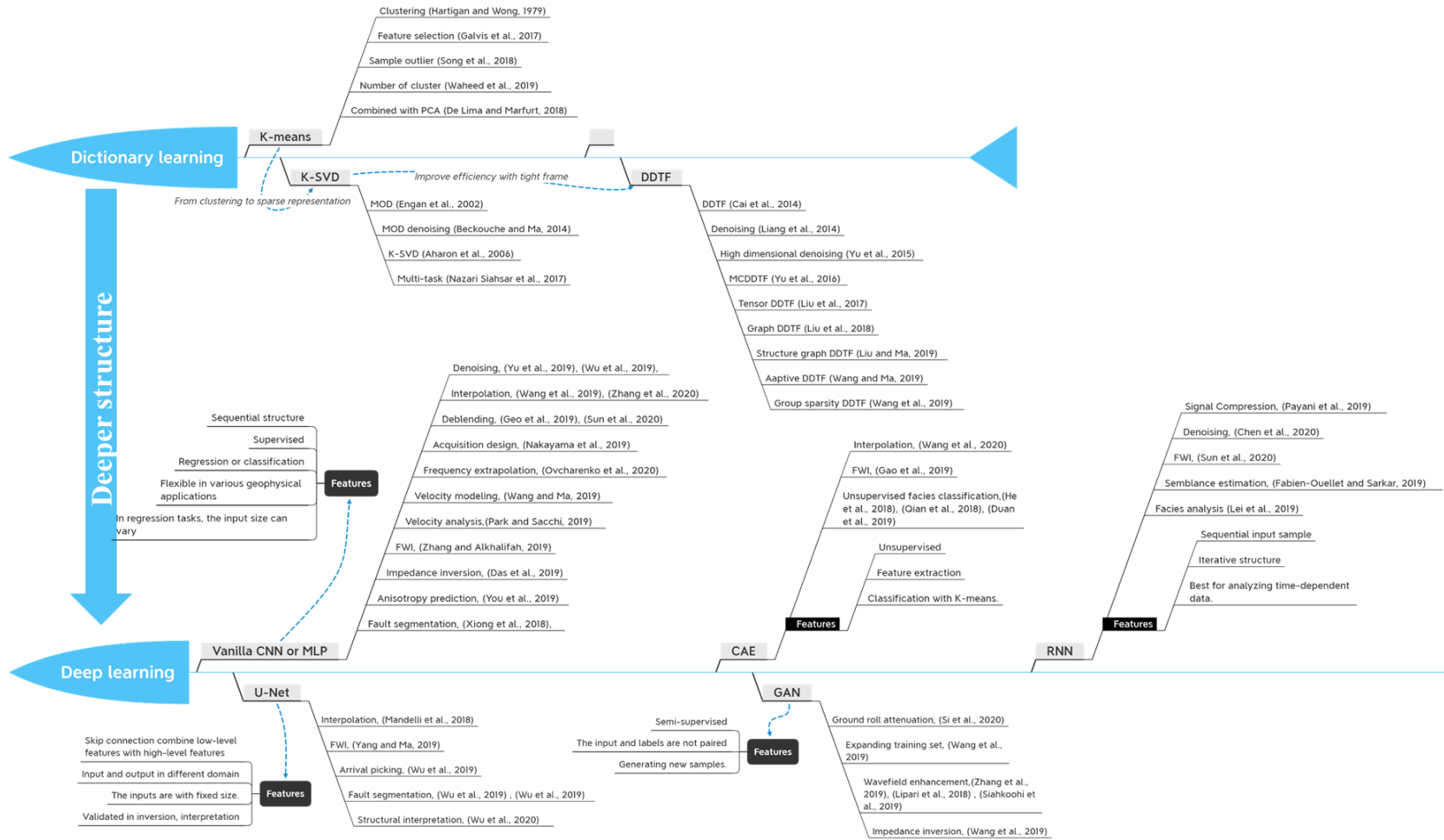

Figure 3. Conceptual map of the data-driven methods in exploration geophysics included in this paper

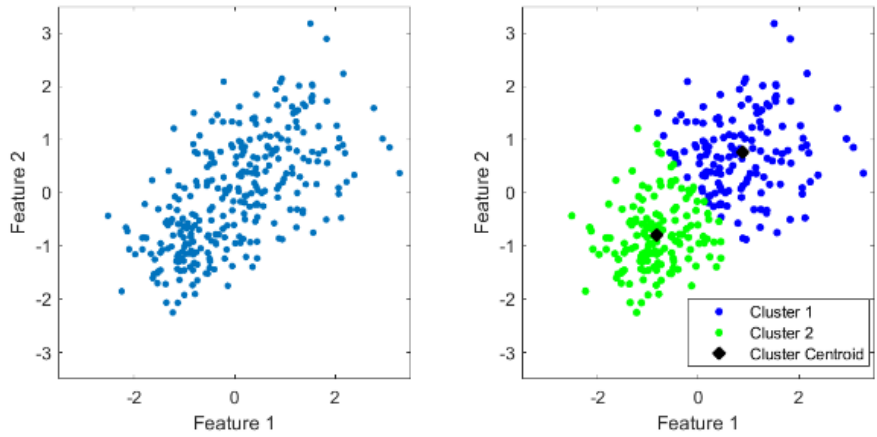

Figure 4. Illustration of the K-means method. Left: A randomly generated dataset with 300 samples (*N*=300) and two features (*M*=2). Right: The classification result (*K*=2).

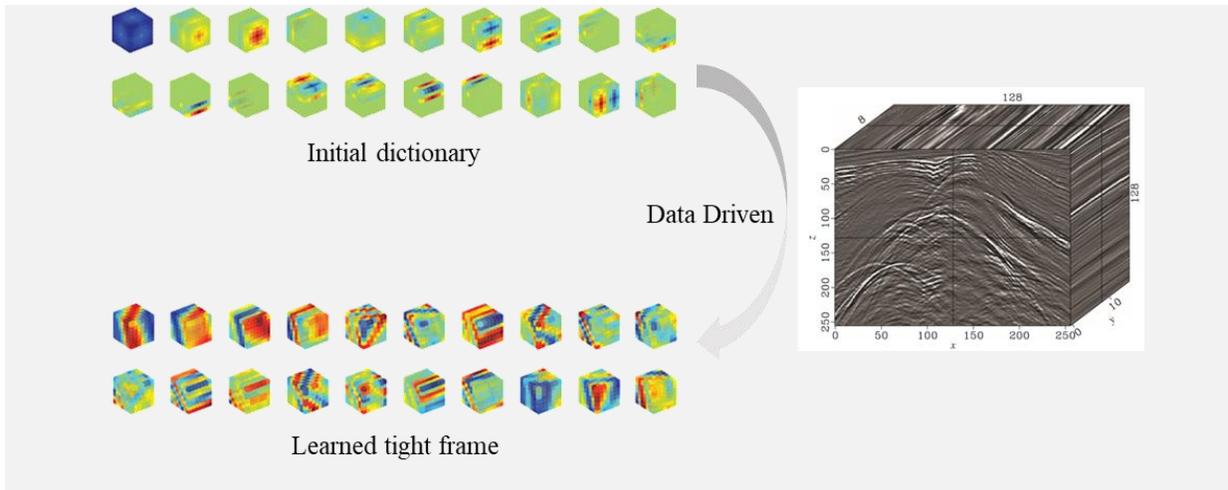

Figure 5. An illustration of DDTF. The dictionary is initialized with a spline framelet. After training based on a post-stack seismic dataset, the trained dictionary exhibits apparent structures.

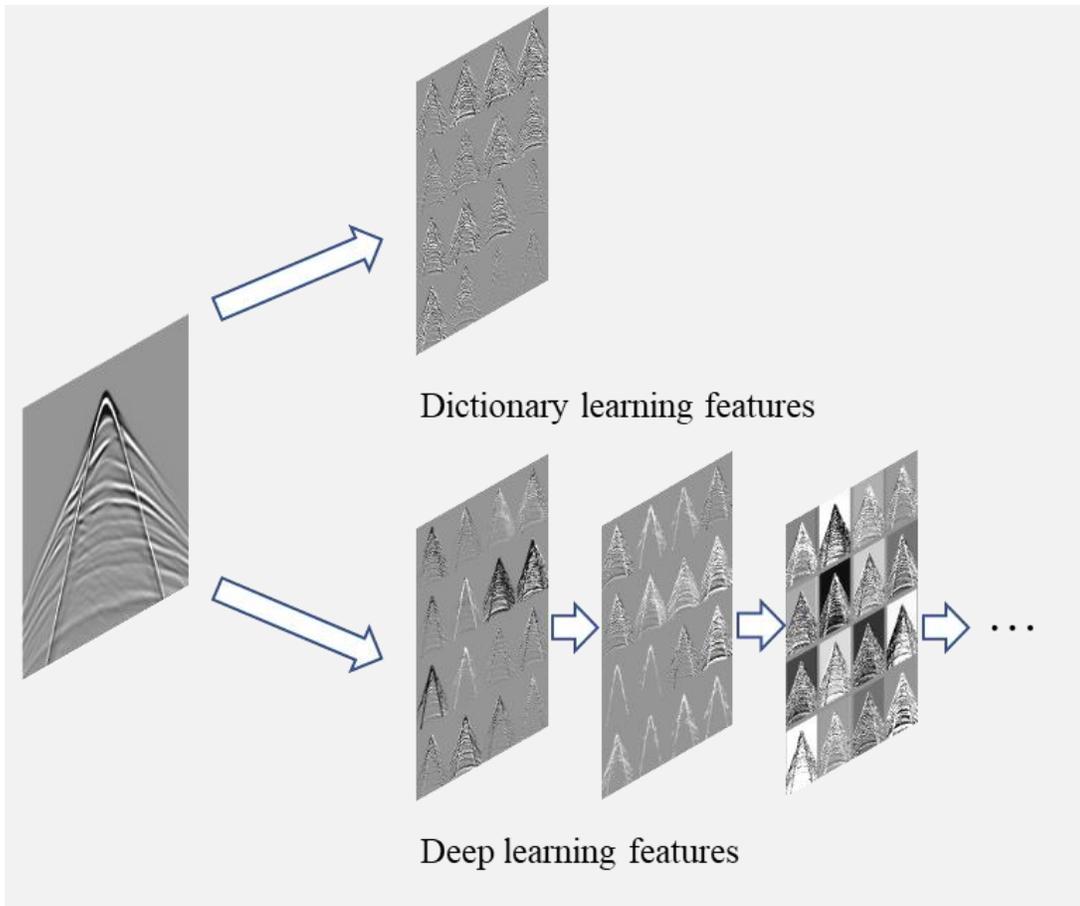

Figure 6. Comparison of the learned features in dictionary learning and deep learning. Dictionary learning obtains single-level decomposed features. Deep learning captures multilevel decomposed features.

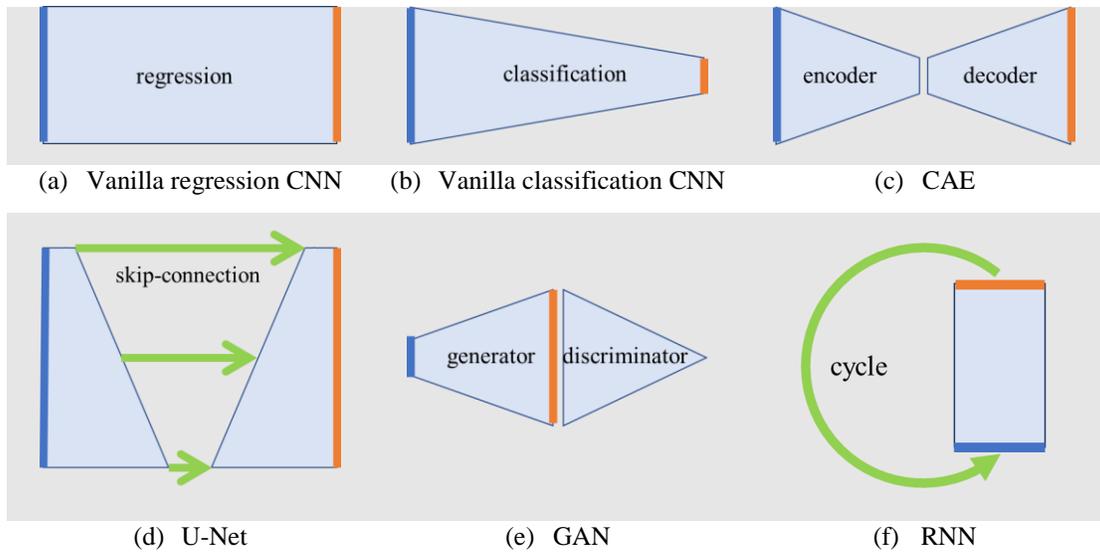

(a)  Vanilla regression CNN    (b)  Vanilla classification CNN    (c)  CAE

(d)  U-Net    (e)  GAN    (f)  RNN

Figure 7. Sketches of DNNs. We omit the details of the layers and maintain the shape of each network architecture. The blue lines indicate inputs, and the orange lines indicate outputs, the length of which represents the data dimension. The green lines indicate intermedia connections. Convolutional layers usually have the same size as the input and output. Pooling layers will reduce the data set size. Unpooling layers will expand the data set size. (a) In regression tasks, such as denoising or interpolation, the output often has the same dimension as the input. (b) In classification tasks, the outputs are labels with a relatively small dimension. (c) The dimension of the latent feature space in the CAE is lower than that of the data space. (d) Skip connections are used to bring the low-level features to a high level in U-Net. (e) In a GAN, low-dimensional random vectors are used to generate a sample from the generator, and then the sample is classified as true or false by the discriminator. (f) In an RNN, the output of the network is used as input in a cycle.

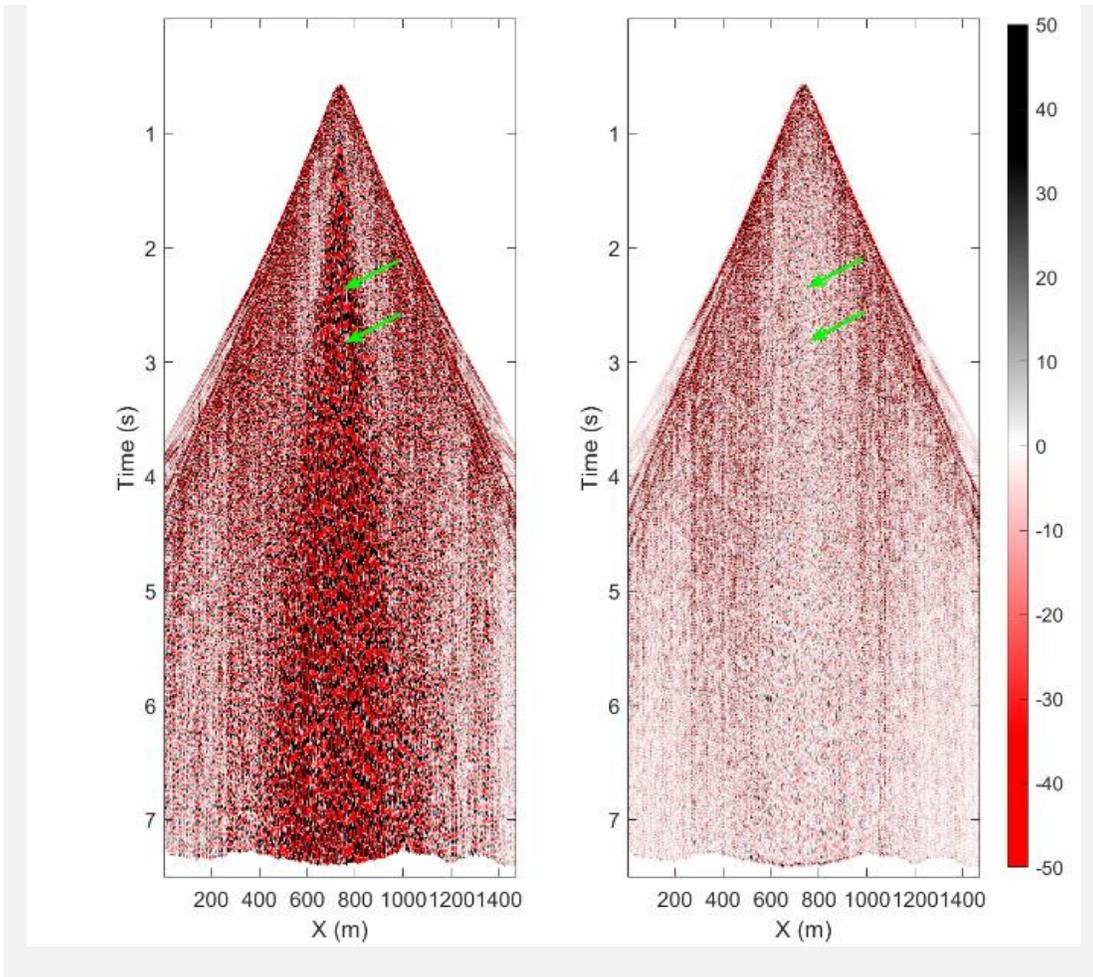

Figure 8. Deep learning for scattered ground-roll attenuation. On the left is the original noisy dataset. On the right is the denoised dataset. The scattered ground roll marked by the green arrows are removed.

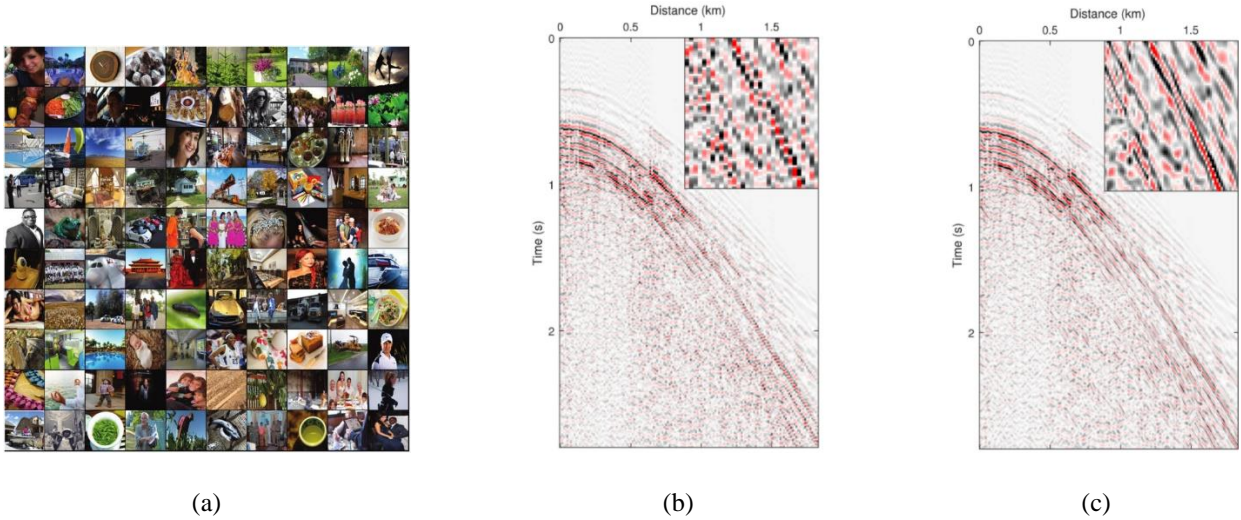

(a)                     (b)                     (c)

Figure 9. The training set and seismic interpolation result (Zhang, Yang et al. 2020). (a) A subset of the natural image dataset. The natural image dataset was used to train a network for seismic data interpolation. (b) An under-sampled seismic record. (c) The interpolated record corresponding to (b). The regions 1.6-1.88 s and 1.0-1.375 km are enlarged at the top-right corner.

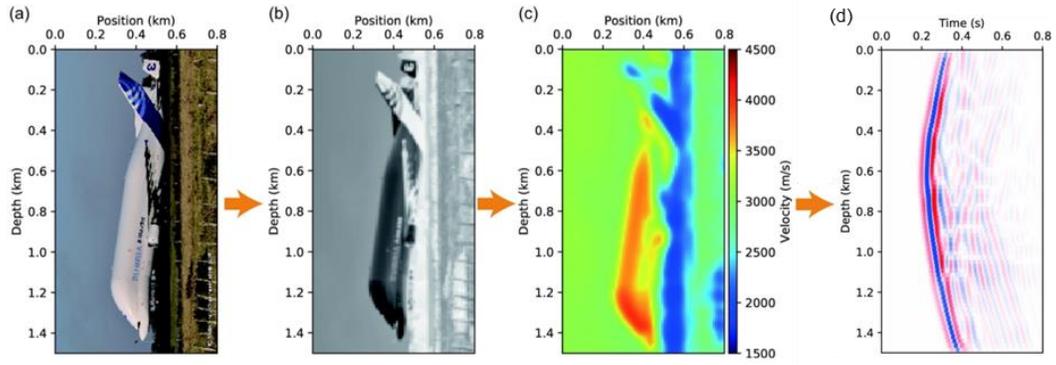

Figure 10. Converting a three-channel color image into a velocity model (Wang and Ma 2020). (a)-(c) are original color image, gray scale image, and corresponding velocity model. (d) is the seismic record generated from a cross-well geometry on (c).

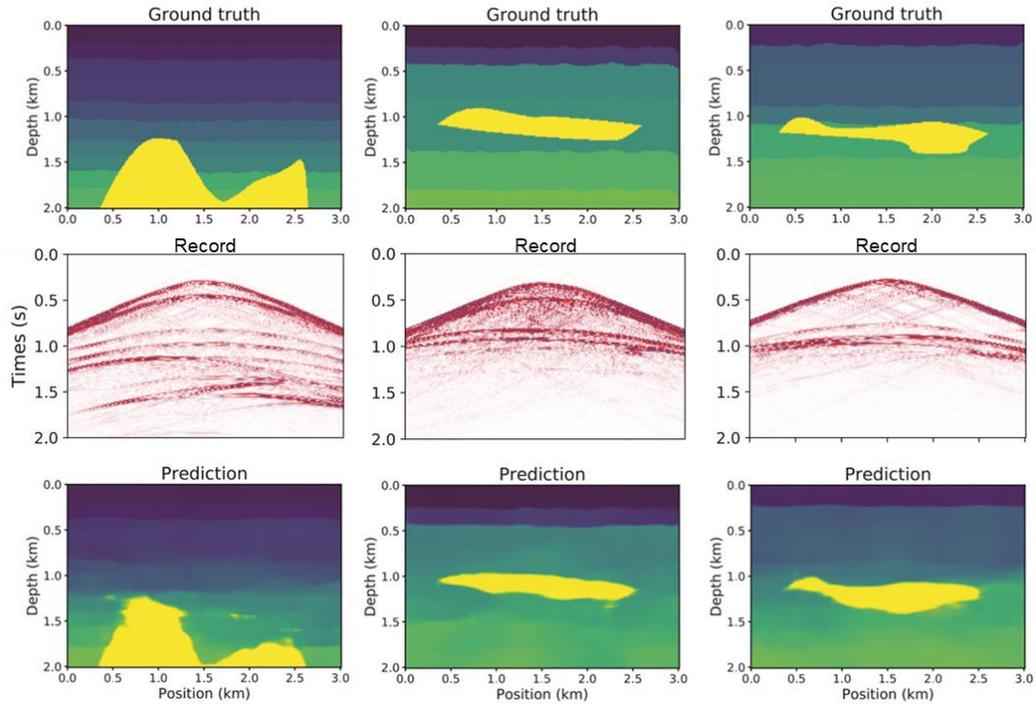

Figure 11. Predicting the velocity model with U-Net from raw seismological data (Yang and Ma 2019). The columns indicate different velocity models. From top to bottom are the ground truth velocity models, generated seismic records from one shot, and the predicted velocity models.

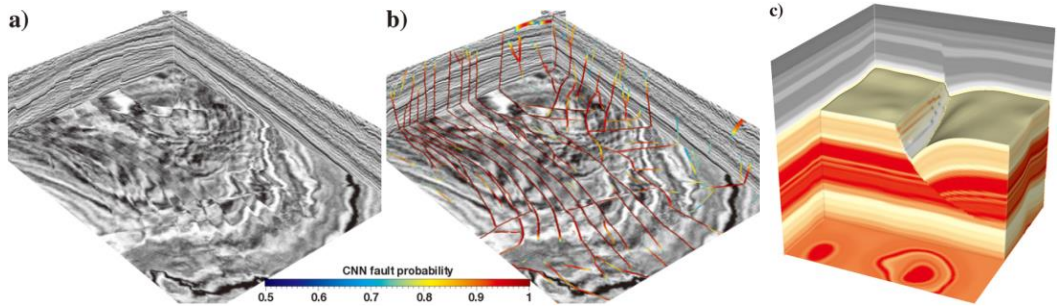

Figure 12. (a) A post-stack dataset. (b) Prediction result of (a). (c) A synthetic dataset (Wu, Geng et al. 2020).

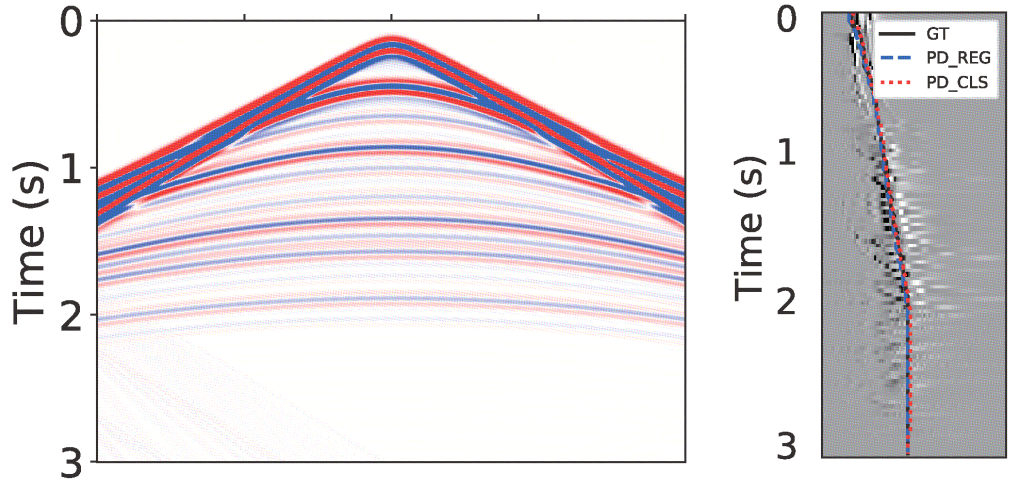

Figure 13. Velocity picking based on U-Net. The inputs are seismological data on the left. The outputs are the picking positions on the right. GT means ground truth. PD_REG and PD_CLS represent the predictions of the regression network and classification network, respectively.

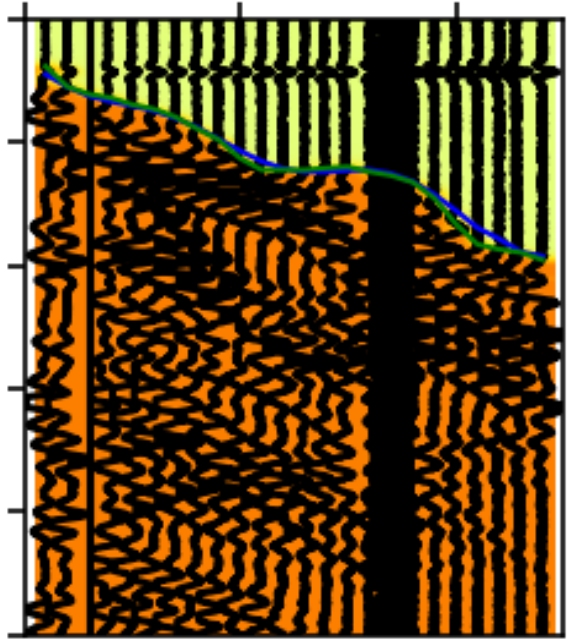

Figure 14. Phase picking based on U-Net. The inputs are seismological data. The outputs are zeros above the first arrival in the green area, ones below the first arrival in the yellow area, and twos for the first arrival on the blue line. The green line indicates the predicted first arrival. This experiment was performed based on the modified code from https://github.com/DaloroAT/first_break_picking.

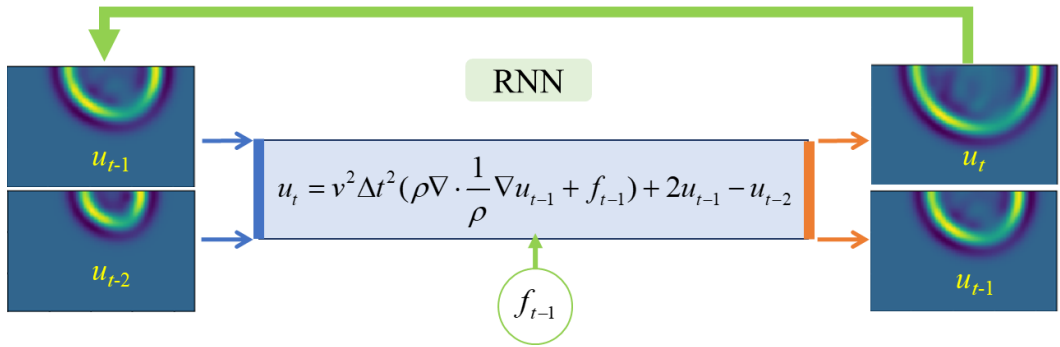

Figure 15. Modified RNN based on the acoustic wave equation for wave modeling (Liu 2020). The diagram represents the discretized wave equation implemented in an RNN. The auto-differential mechanics of a DNN help to efficiently optimize the velocity and density.

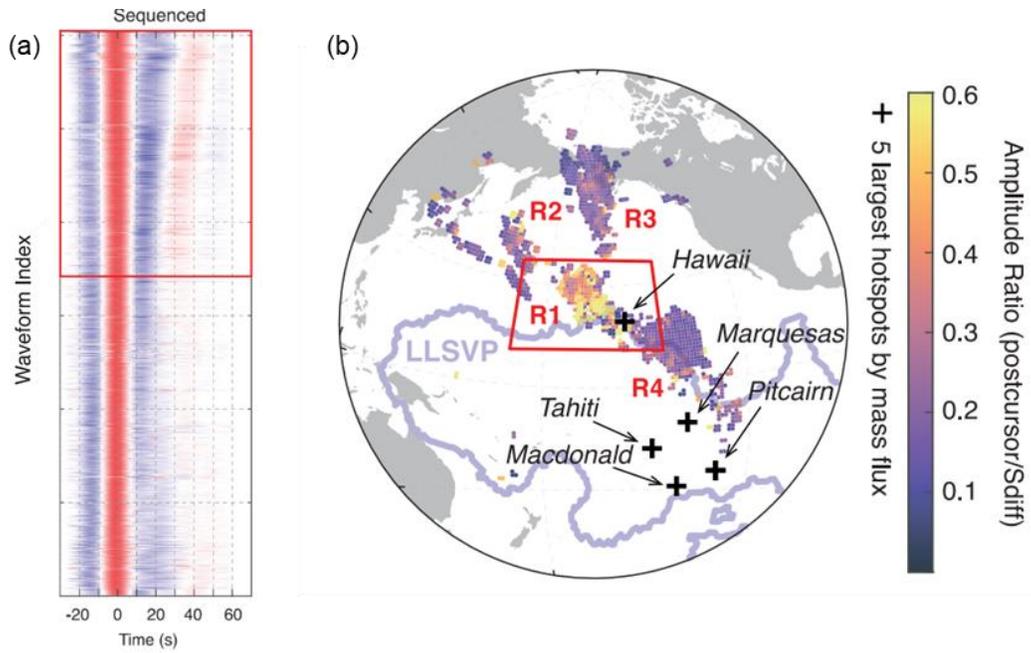

Figure 16 (a) Sequencer ordering enables the identification of a substantial (~40% of all waveforms) subpopulation of Sdiff postcursors (red box). (b) Stacks of postcursor amplitude relative to main Sdiff arrival averaged in 1° bins (Kim et al. 2020).

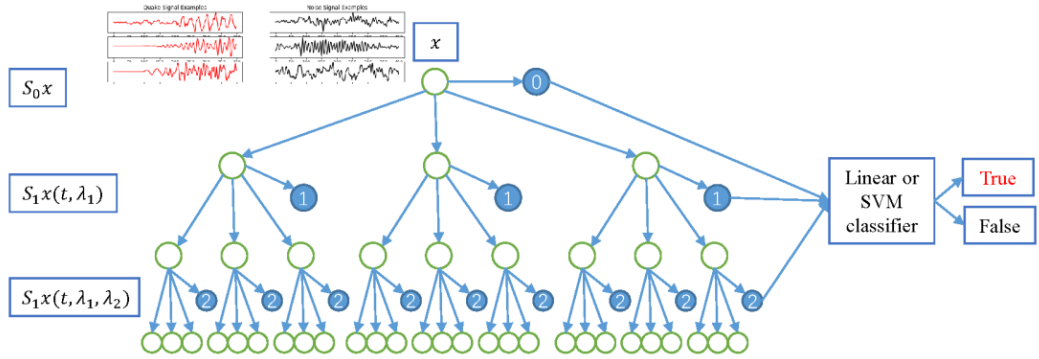

Figure 17. (a) The architecture of WST. Unlike in a CNN, the outputs of WST are combined with the outputs of each layer. Then, the outputs of WST serve as features for a classifier.

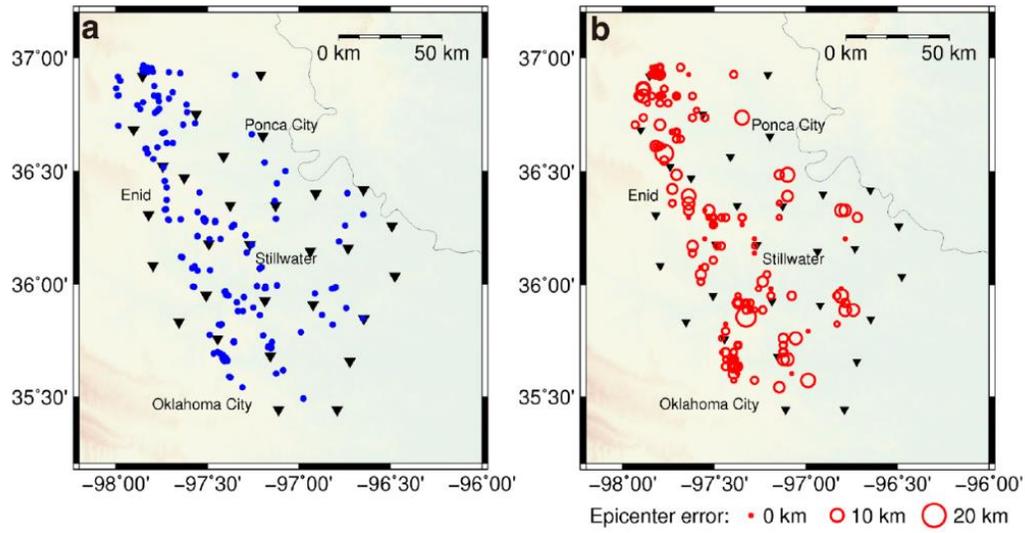

Figure 18. Locating earthquake sources with deep learning. The black triangles are stations. Left: the blue dots are the actual locations. Right: the red circles are the predicted locations. The radius of a circle represents the predicted epicenter error (Zhang et al. 2020).

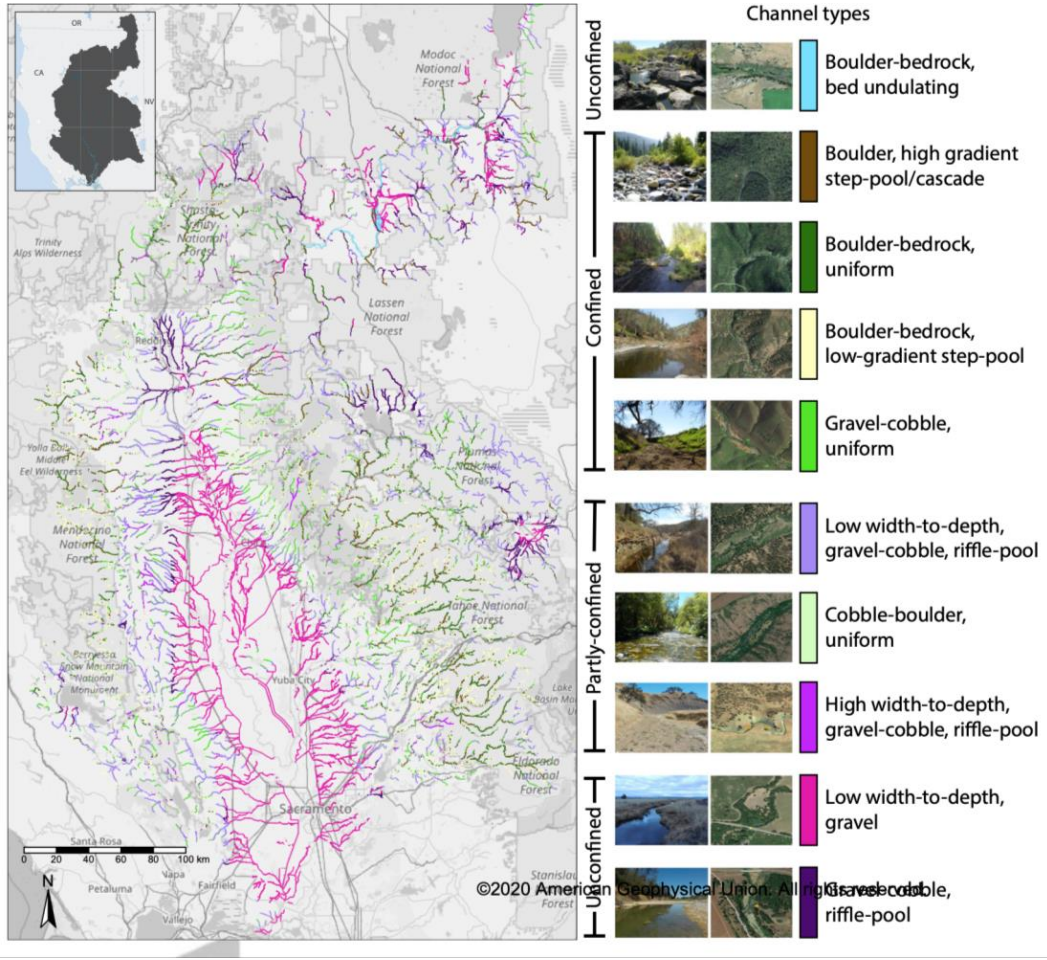

Figure 19. Machine learning predictions of stream channel type for the Sacramento Basin in California, USA (Guillon et al. 2020).

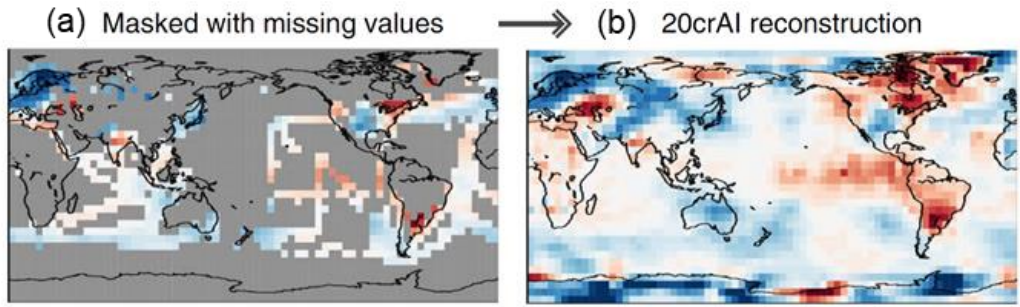

Figure 20 AI models reconstruct temperature anomalies with many missing values (Kadow et al. 2020).

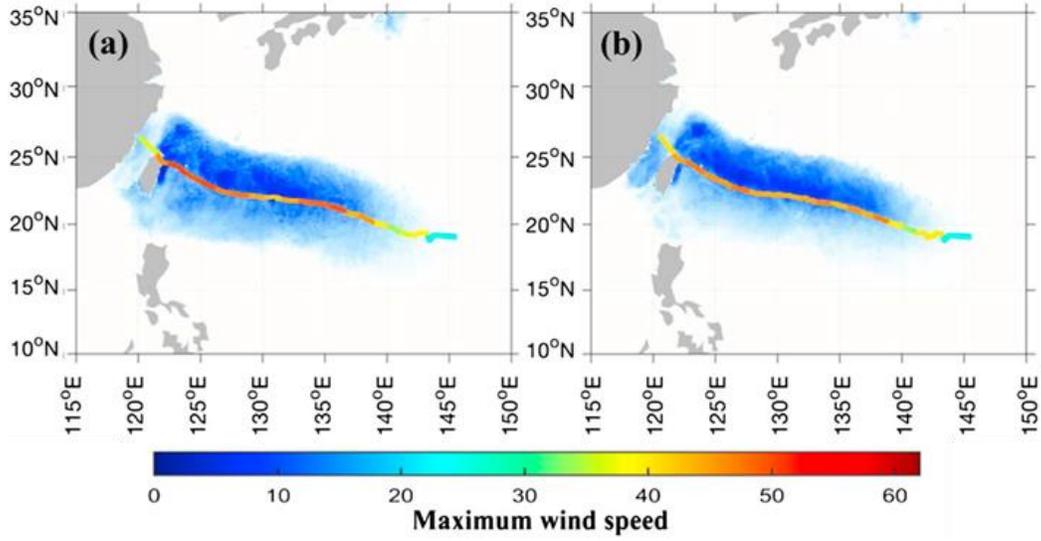

Figure 21 Typhoon track prediction with (a) shallow learning algorithm and (b) deep learning algorithm (Jiang et al. 2018).

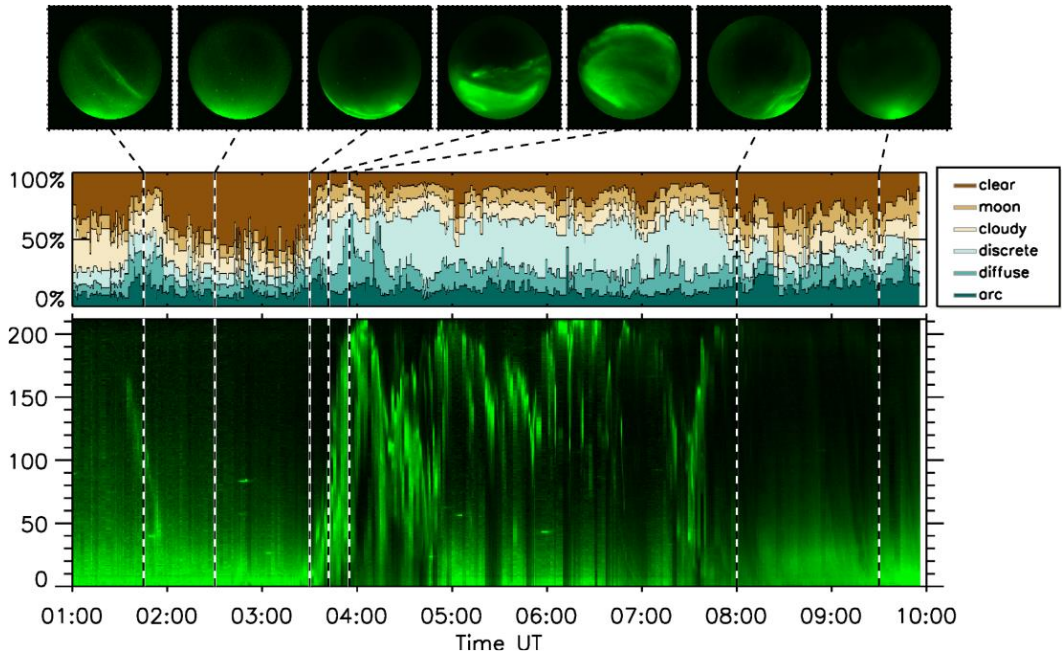

Figure 22 The bottom panel shows a keogram from auroral data collected on 21 January 2006 at Rankin Inlet. The keogram consists of single column from the auroral images at different time. The middle panel shows the probabilities for the six categories as predicted by the ridge classiffier trained with the entire training dataset. At the top are auroral images at different times. (Clausen and Nickisch 2018)

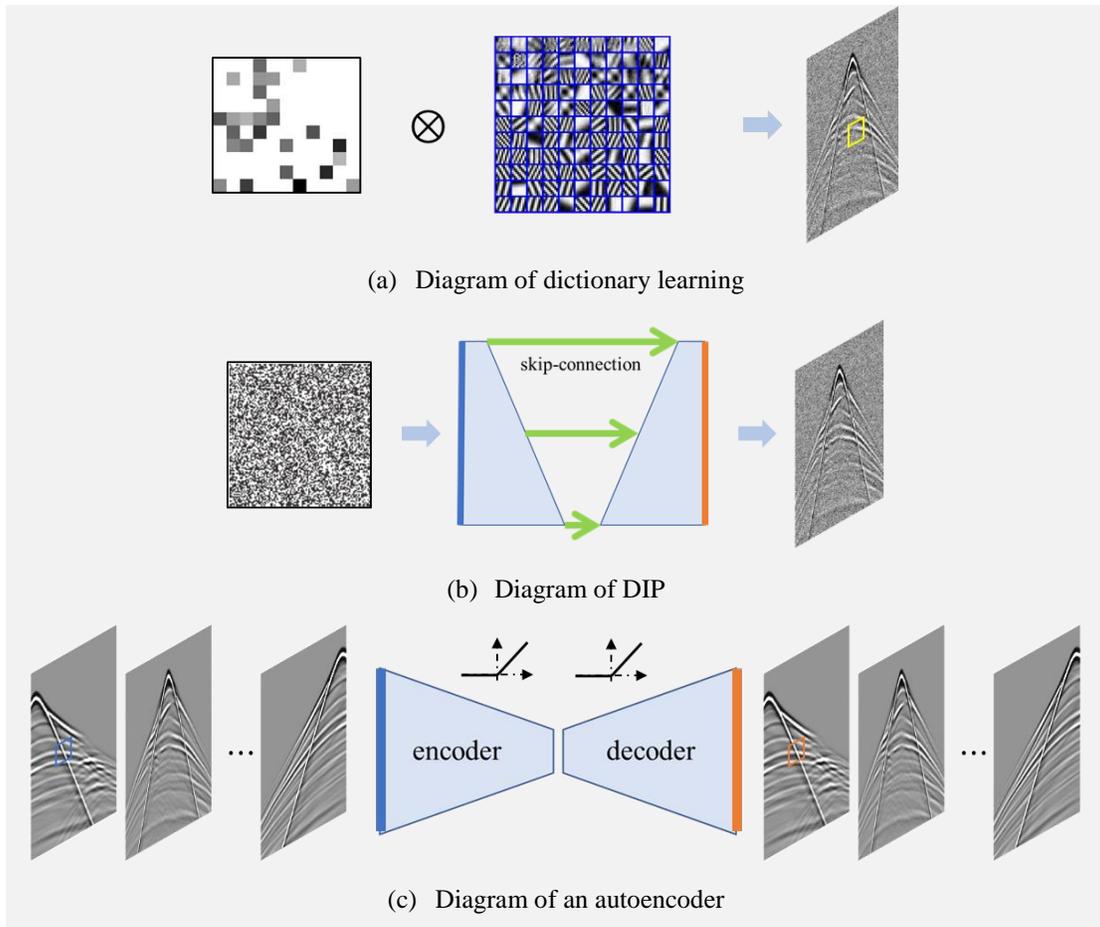

(a) Diagram of dictionary learning

(b) Diagram of DIP

(c) Diagram of an autoencoder

Figure 23. Comparison of dictionary learning and deep learning. (a) Dictionary learning has a shallow structure and is unsupervised and linear. A sparsity constraint is placed on the coefficients. (b) In DIP, the network architecture constrains the produced image. (c) An autoencoder has a deep structure, an extensive training set, and nonlinear operators.

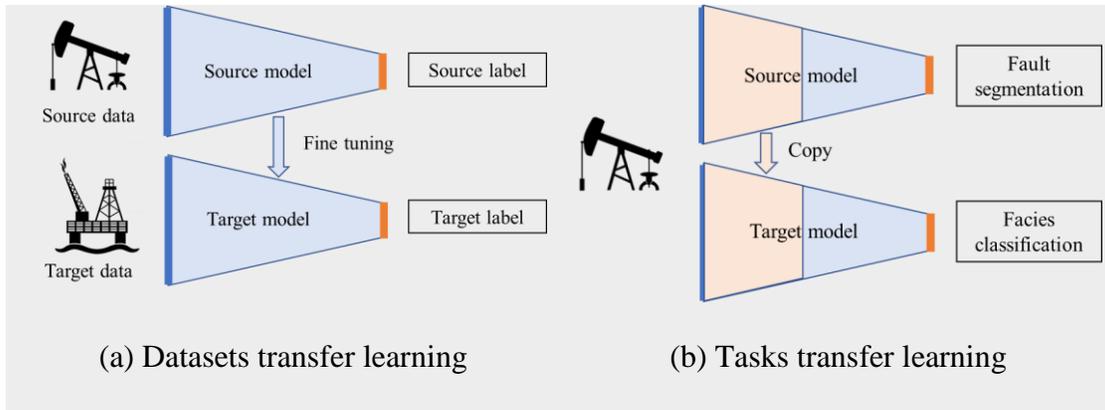

(a) Datasets transfer learning         (b) Tasks transfer learning

Figure 24. Diagrams of transfer learning. (a) Transfer learning between different datasets. The parameters of one trained model can be moved to another model as initialization conditions. (b) Transfer learning between different tasks. The first layers of one trained model can be copied to another model.

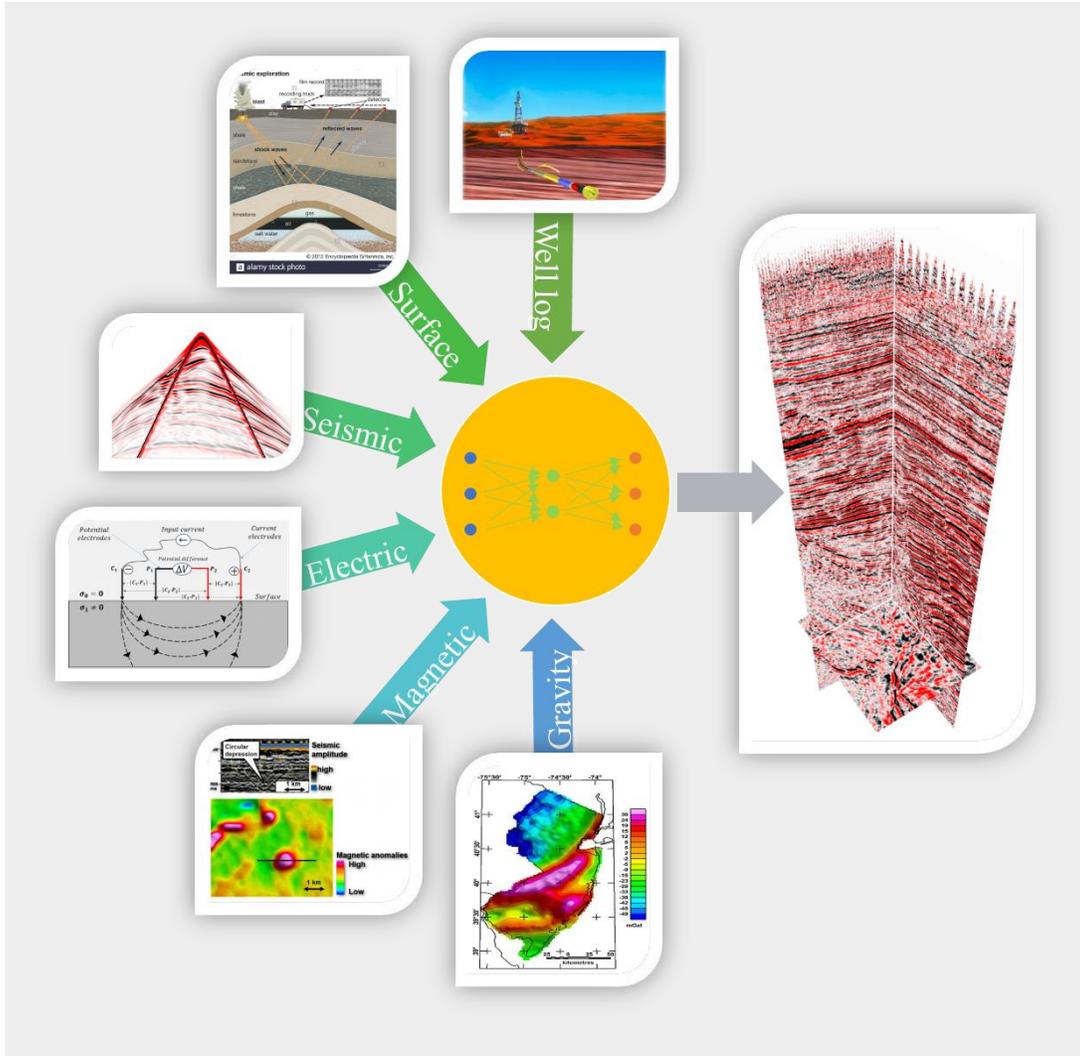

Figure 25. An illustration of multimodal deep learning

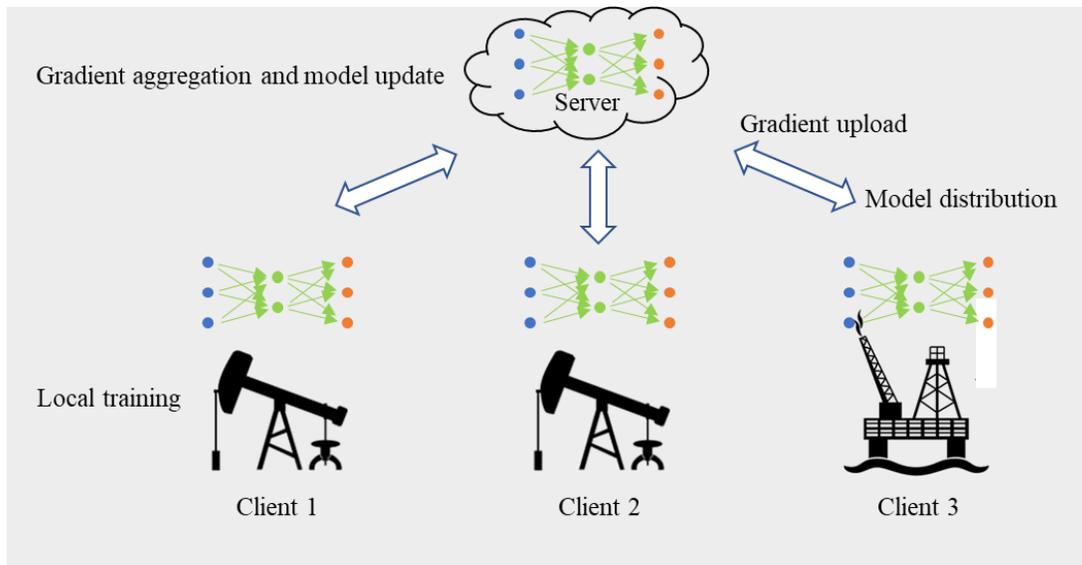

Figure 26. Federated learning. The clients train the DNN with local datasets and uploads the model gradient to the server. The server aggregates the gradients and updates the global model. Then, the updated model is distributed to all the local clients. Many rounds of training are performed until the model meets a certain accuracy requirement.

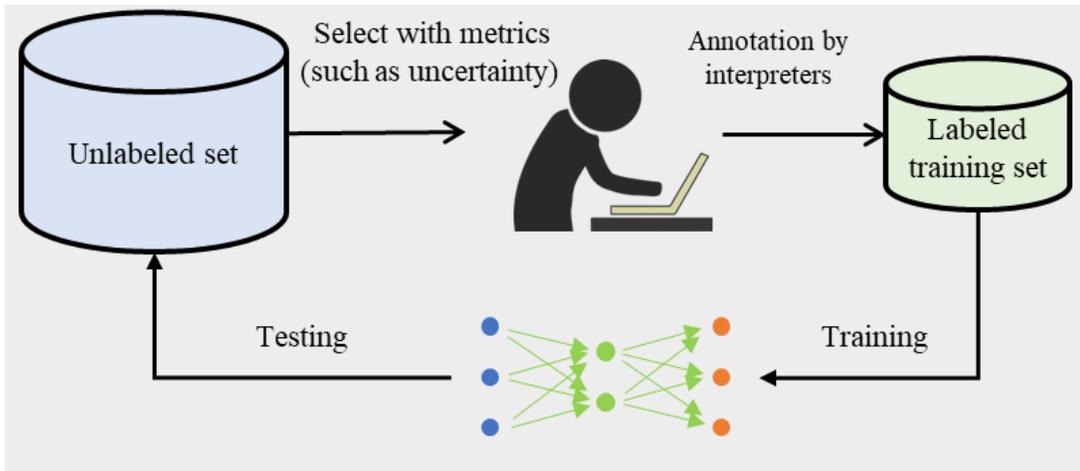

Figure 27. An illustration of active learning. We choose samples with high uncertainty and manually annotate them to serve as training samples.